\documentclass[11pt]{article}
\usepackage[T1]{fontenc}
\usepackage[latin1]{inputenc}
\usepackage{amsmath}
\usepackage{setspace}
\usepackage{amssymb,amsfonts}
\usepackage[english]{babel}

\newcommand{\noun}[1]{\textsc{#1}}

\def\Xint#1{\mathchoice
{\XXint\displaystyle\textstyle{#1}}%
{\XXint\textstyle\scriptstyle{#1}}%
{\XXint\scriptstyle\scriptscriptstyle{#1}}%
{\XXint\scriptscriptstyle\scriptscriptstyle{#1}}%
\!\int}
\def\XXint#1#2#3{{\setbox0=\hbox{$#1{#2#3}{\int}$}
\vcenter{\hbox{$#2#3$}}\kern-.5\wd0}}

\def\dashint{\Xint-}

\begin{document}

\title{\hfill {\small Brown-HET-1469}\\
Notes on  Collective Field Theory of  Matrix and Spin Calogero
Models}

\author{\noun{Inês Aniceto}\footnote{email:nes@het.brown.edu}$\quad$ and $\ $ \noun{Antal Jevicki}\footnote{email:antal@het.brown.edu}\\
 Physics Department\\
Brown University\\
Providence, Rhode Island 02912, USA \\
}

\maketitle

\begin{abstract}
Matrix models and related Spin-Calogero-Sutherland models are of
major relevance in a variety of subjects, ranging from condensed
matter physics to QCD and low dimensional  string theory. They are
characterized by integrability and exact solvability. Their
continuum, field theoretic representations are likewise of definite
interest. In this paper we describe various continuum, field
theoretic representations of these models based on bosonization and
collective field theory techniques. We  compare various known
representations and describe some nontrivial applications.

\end{abstract}

\bigskip \thispagestyle{empty} \hfill

\vskip 15mm

\bigskip

\vskip10mm

\parskip=7pt

\newpage

\section{Introduction}

Matrix models in general and specially their simpler N-body cousins
can be studied effectively in the large N limit through continuum,
collective field type techniques. The one matrix problem in its
fermionic reduction gave a systematic perturbative representation of
low  dimensional noncritical string theory
\cite{Klebanov:1991qa,Jevicki:1993qn,Ginsparg:1993is,Alexandrov:2003ut,Das:1990ka}.
Its continuum, collective field representation gave major insight
into  the origin of the extra (Liouville) dimension and the tachyon
mode. On the other hand, understanding of nonperturbative effects
such as vortices and black holes clearly requires the full
understanding of the theory
\cite{Gross:1990md,Boulatov:1991xz,Kazakov:2000pm}, including
possibly gauge degrees of freedom. Recently, for example the
dynamics of long strings or FZZT branes was successfully simulated
by nonsinglet degrees of the matrix \cite{Maldacena:2005hi}.
 The continuum, field theoretic
representations of these theories can be formulated through techniques
of non-abelian bosonization or collective field theory
\cite{Uglov:1996ba,Avan:1995sp,Avan:1996vi,Awata:1995by}. The
resulting hamiltonians are written in conformal field theory
notation and exhibit higher, Yangian symmetry of the form studied
in
\cite{Bernard:1992ya,Babelon:1991ah,Haldane:1992sj,Bernard:1993va,Billey:1994tv,Schoutens:1994au}.

Because of their relevance it is of considerable interest to study
further these continuum field theoretic  representations. They hold
the potential for giving answers to a number of (nonperturbative)
problems both in string theory and condensed matter physics.

In this  review we describe in some detail the continuum, collective
field theory techniques and study the relationship between the
continuum field  theories that result. We also demonstrate the
usefulness of the bosonic, collective field representation by
presenting some nontrivial applications: in one we discuss in the
field theory setting the linearized spectrum equation, which is
shown to  take the form of an eigenvalue problem due to Marchesini
and Onofri \cite{Marchesini:1979yq}. This eigenvalue problem has
recently appeared in number of matrix theory applications. Our
second demonstration concerns the treatment of nontrivial
backgrounds associated with nontrivial states of the theory. In
particular we study the theory in  a semiclassical background
associated with the quantum many-body state found by Haldane and Ha
\cite{ha:1992-}. This gives a highly nontrivial example where exact
results are known and serves as a demonstration  of how nontrivial
backgrounds appear in the collective field framework.

The outline of the review is as follows. In section
\ref{Introductory Section} we introduce the spin Calogero-Sutherland
model and describe in detail its relationship with the  matrix
model. In section \ref{Current-Algebra Rep} we give details of the
bosonization procedure of the spin CS hamiltonian in the conformal
field theory language. In section \ref{Collective FT} the same
hamiltonian is studied using collective fields. Sections \ref{M-O
Application} and \ref{Nontrivial Bckgrd} describe some non trivial
applications of this theory. Possible, further applications are also
discussed in the Conclusion.

\section{From Matrix Models to Spin Calogero-Sutherland
Models}\label{Introductory Section}

Various reductions of matrix models lead to simpler many-body
problems of Spin Calogero-Sutherland type
\cite{Affleck:1981xf,Tan:1981yz,Andric:1982jk,Gibbons:1984,Wojciechowski:1985,Polychronakos:1992zk,Dabholkar:1991te,
Ferretti:1993fu,Ferretti:1990zz,Minahan:1992bz,Minahan:1992ht,Jevicki:1991yk,Rodrigues:1992by,Andric:1994su,
Andric:1996nn,Andric:1998hi,Sen:1997qt}. We begin by describing the
general scheme for such reductions (in the case of Euclidean metric,
the more general case of Riemannian symmetric spaces can be seen in
recent work \cite{Feher:2005pe,Feher:2006qt}). Consider matrix
quantum mechanics defined by the hamiltonian
\cite{Marchesini:1979yq,Boulatov:1991xz}\begin{equation}
H=\rm{Tr}\left(\frac{1}{2}\Pi^{2}+\left(\frac{N}{g}\right)V\left(\left(\frac{g}{N}\right)\Phi^{2}\right)\right).\label{Matrix
Hamiltonian}\end{equation}
 Here $\Phi$ represents a Hermitian $N\times N$ matrix and the scaling of coupling constants (by $N$) in the general,
 polynomial potential $V$ is done in accordance with 't Hooft's large $N$ limit.
 The hamiltonian acts on the Hilbert space of square integrable functions defined
with respect to the invariant measure
$\left[d\Phi\right]=\prod\Phi_{aa}\prod_{a<b} d\Phi_{ab}d\Phi_{ba}$.

The invariance of the theory under $SU\left(N\right)$
transformations implies that the eigenstates are described by
unitary irreducible representations of $SU\left(N\right)$. To
exhibit the group theoretic degrees of freedom it is convenient to
use a {}``polar'' decomposition of the matrix coordinate $\Phi$, and
perform a separation of eigenvalues (of the matrix) and the
angles\begin{equation}
\Phi=\Omega^{\dagger}\Lambda\Omega\qquad\Leftrightarrow\qquad\Phi_{ab}=\sum_{i}\left(\Omega^{\dagger}\right)_{ai}\lambda_{i}\left(\Omega\right)_{ib}\,,\label{Diagonalization
of matrix}\end{equation}
 where $\Lambda=\rm{diag}\left(\lambda_{1},...,\lambda_{N}\right)$.
 Indices
$a,b,c=1,...,N$, and $i,j=1,...,N$, will denote the internal indices
throughout. One notes that $\Omega\in SU\left(N\right)/\mathbb{H}$,
with $\mathbb{H}$ being the stability subgroup of $\Lambda$ under
$SU\left(N\right)$. Equation (\ref{Diagonalization of matrix}) can
be rewritten as \begin{equation}
\Phi_{ab}\left(\Omega^{\dagger}\right)_{bi}=\lambda_{i}\left(\Omega^{\dagger}\right)_{ai},\end{equation}
 so we can conclude that $\left(\Omega^{\dagger}\right)_{bi}$ is
the component $b$ of the $i$th eigenvector of the matrix $\Phi$,
with corresponding eigenvalue $\lambda_{i}$.

Defining vectors $Z^{i}$ and $\bar{Z}^{i}$ by their components $Z_{\, a}^{i}=\Omega_{ia}$
and $\bar{Z}_{a}^{\, i}=\left(\Omega^{\dagger}\right)_{ai}=\bar{\Omega}_{ia}$
($\bar{Z}_{a}^{\, i}$ is the complex conjugate of $Z_{\, a}^{i}$),
we can determine the constraints on these vectors coming from $\Omega\in U\left(N\right)$.
Because $\Omega$ is unitary, we have\begin{eqnarray*}
\Omega^{\dagger}\Omega=1\!\!\rm{l} & \quad\Rightarrow\quad & \sum_{i}\bar{Z}_{a}^{\, i}Z_{\, b}^{i}=\delta_{ab},\\
\Omega\Omega^{\dagger}=1\!\!\rm{l} & \quad\Rightarrow\quad &
\sum_{a}Z_{\, a}^{i}\bar{Z}_{a}^{\, j}=\delta_{ij}.\end{eqnarray*}
 These two constraints state the completeness and orthonormality of
the eigenfunctions, respectively. We can write the eigenfunction equation
in terms of the eigenvectors $\bar{Z}^{i}$ as:\begin{equation}
\Phi_{ab}\bar{Z}_{b}^{\, i}=\lambda^{i}\bar{Z}_{a}^{\, i}.\end{equation}

To evaluate the form of the hamiltonian after the above
transformation one needs the following identities, obtained by using
the chain rule:\begin{eqnarray*}
\frac{\delta\lambda^{i}}{\delta\Phi_{bc}} & = & Z_{\, b}^{i}\bar{Z}_{c}^{\, i},\\
\frac{\delta\bar{Z}_{a}^{\, i}}{\delta\Phi_{bc}} & = & \sum_{j\left(\ne i\right)}\frac{\bar{Z}_{a}^{\, j}Z_{\, b}^{j}\bar{Z}_{c}^{\, i}}{\lambda_{i}-\lambda_{j}},\\
\frac{\delta Z_{\, a}^{i}}{\delta\Phi_{bc}} & = & \sum_{j\left(\ne i\right)}\frac{Z_{\, b}^{i}\bar{Z}_{c}^{\, j}Z_{\, a}^{j}}{\lambda_{i}-\lambda_{j}}.\end{eqnarray*}
 We can now apply these rules to the kinetic term (we will use the
notation where repeated indices are summed over, unless stated otherwise).
First:\begin{eqnarray*}
\frac{\partial}{\partial\Phi_{bc}} & = & Z_{\, b}^{i}\bar{Z}_{c}^{\, i}\frac{\partial}{\partial\lambda_{i}}+\sum_{j\left(\ne i\right)}\frac{Z_{\, b}^{i}\bar{Z}_{c}^{\, j}Z_{\, a}^{j}}{\lambda_{i}-\lambda_{j}}\frac{\partial}{\partial Z_{a}^{i}}+\sum_{j\left(\ne i\right)}\frac{\bar{Z}_{a}^{\, j}Z_{\, b}^{j}\bar{Z}_{c}^{\, i}}{\lambda_{i}-\lambda_{j}}\frac{\partial}{\partial\bar{Z}_{a}^{\, i}},\end{eqnarray*}
 and then:\begin{eqnarray}
-\rm{Tr}\left(\Pi^{2}\right) & = &
\sum_{b,c}\frac{\partial}{\partial\Phi_{cb}}\frac{\partial}{\partial\Phi_{bc}}
  =  \sum_{i}\frac{\partial^{2}}{\partial\lambda_{i}^{2}}+2\sum_{i\ne j}\frac{1}{\lambda_{i}-\lambda_{j}}\frac{\partial}{\partial\lambda_{i}}+\sum_{i\ne j}\frac{\bar{Q}_{ij}Q_{ij}}{\left(\lambda_{i}-\lambda_{j}\right)^{2}},\end{eqnarray}
 where we defined \begin{eqnarray}
Q_{ij} & \equiv & i\sum_{a}\left\{ Z_{\, a}^{i}\frac{\partial}{\partial Z_{\, a}^{j}}-\bar{Z}_{a}^{\, j}\frac{\partial}{\partial\bar{Z}_{a}^{\, i}}\right\} \equiv i\left\{ Z^{i}\frac{\partial}{\partial Z^{j}}-\bar{Z}^{j}\frac{\partial}{\partial\bar{Z}^{i}}\right\} ,\\
\bar{Q}_{ij} & = & Q_{ji}=i\left\{ Z^{j}\frac{\partial}{\partial Z^{i}}-\bar{Z}^{i}\frac{\partial}{\partial\bar{Z}^{j}}\right\} .\end{eqnarray}
 The sum in the matrix variable $a$ has been omitted. The $Q_{ij}$
is a set of noncommuting differential operators that only depend on
the angular coordinates. An equivalent way of writing the result
above is \cite{Marchesini:1979yq}:\begin{equation}
\rm{Tr}\left(\Pi^{2}\right)=-\frac{1}{\Delta}\sum_{i}\frac{\partial^{2}}{\partial\lambda_{i}^{2}}\Delta-\sum_{i\ne
j}\frac{Q_{ij}Q_{ji}}{\left(\lambda_{i}-\lambda_{j}\right)^{2}}\,.\label{Hamiltonian
separated in eigenvalues and angles}\end{equation}
 Here $\Delta=\prod_{i<j}\left(\lambda_{i}-\lambda_{j}\right)$ is
the Vandermonde determinant, representing the invariant measure
$\left[d\Phi\right]=\Delta^{2}d\lambda_{1}\cdots
d\lambda_{N}\left[d\Omega\right]$, where $\left[d\Omega\right]$ is
the invariant volume element in $U\left(N\right)$. An equivalent way
of deriving (\ref{Hamiltonian separated in eigenvalues and angles}),
based on the line element, can be found for example in
\cite{Marchesini:1979yq}. In general, the potential terms are easily
expressible in terms of the eigenvalues and the full Matrix
hamiltonian becomes\begin{eqnarray}  H & = &
-\frac{1}{2}\frac{1}{\Delta}\sum_{i}\frac{\partial^{2}}{\partial\lambda_{i}^{2}}\Delta+\frac{N}{g}\sum_{i}V\left(\left(\frac{g}{N}\right)\lambda_{i}^{2}\right)+\sum_{i\ne
j}\frac{Q_{ij}^{\mathcal{R}}Q_{ji}^{\mathcal{R}}}{\left(\lambda_{i}-\lambda_{j}\right)^{2}}\,.\label{Effective
Matrix Hamiltonian}\end{eqnarray}

 We have put in this last result the explicit dependence
 on the representation $\mathcal{R}$ of $SU\left(N\right)$, as
the $Q_{ij}^{\mathcal{R}}$ are the matrices corresponding to the
$ij$ generator of $SU\left(N\right)$ in the representation
$\mathcal{R}$ \cite{Maldacena:2005hi,Boulatov:1991xz}, since the
theory can be studied in each sector separately. \footnote{One
further comment is in place. The possible representations
$\mathcal{R}$ of $SU\left(N\right)$ are restricted to irreducible
representations which have a state with all weights equal to zero
\cite{Gross:1990md,Maldacena:2005hi}.}
 In the simplest case of $SU\left(N\right)$
invariant \textbf{singlet states} the matrix model reduces to a
many-body problem of free fermions. The system with $V$ being an
inverted oscillator represents 2D non-critical string theory.
Classical hamiltonian reduction with a particular (non zero) value
of angular momentum is known to lead to a Calogero-type hamiltonian
for the eigenvalues \cite{Calogero:1975ii,Moser:1976}. In order to
obtain a trigonometric form of the hamiltonian, namely the
Calogero-Sutherland Model, one uses unitary matrix model (with
$\Phi$ unitary). Continuing our discussion at the quantum level the
next relevant case is that of \textbf{adjoint states} of
$SU\left(N\right)$, studied first in \cite{Marchesini:1979yq} and a
number of works thereafter.

 Models of similar type also show up in the singlet sector of
matrix-vector type theories \cite{Avan:1996vi}. We have in the above
introduced a notation where the eigenvectors of the matrix appear as
vectorlike variables to emphasize the correspondence with that type
of theories. In the full matrix theory case the  flavour index a
clearly ranges from 1 to N . One can then consider a reduction in
the number of vector degrees of freedom to a fixed number R .
 The particular case of a spin chain appears if
we represent the operators of the SU(N) algebra through fermions
$Q_{ij}=\sum_{\alpha=1}^{R}\Psi_{\alpha}^{\dagger}\left(i\right)\Psi_{\alpha}\left(j\right),$
and constrain the value of the number operator at each site \[
\mathcal{N}_{i}\equiv\sum_{\alpha=1}^{R}\Psi_{\alpha}^{\dagger}\left(i\right)\Psi_{\alpha}\left(i\right)=1.\]
 Here $\Psi_{\alpha}\left(i\right)$ ($\alpha=1,...,R$) are
complex $U\left(N\right)$-vector fermionic degrees of freedom, and
$R$ labels the number of independent components acting like a
flavour index \cite{Affleck:1981xf,Tan:1981yz}.

With the above constraint we can write \[
Q_{ij}Q_{ji}=\left(1-\sum_{\alpha,\beta=1}^{R}Q_{\alpha\beta}\left(i\right)Q_{\beta\alpha}\left(j\right)\right),\]
 by defining
 $Q_{\alpha\beta}\left(i\right)\equiv\Psi_{\alpha}^{\dagger}\left(i\right)\Psi_{\beta}\left(i\right)$,
and obtain a representation with generators at each site:\[
Q_{ij}Q_{ji}=\left(1-J_{i}\cdot J_{j}\right).\]
 This corresponds to a Haldane-Shastry spin-spin interaction
\cite{Gibbons:1984,Wojciechowski:1985,Polychronakos:1992zk,ha:1992-}.

Our main concern in the present review is the continuum, field
theoretic representation of this class of models. This
representation is based on composite, collective fields and can be
reached through various bosonization procedures. The basic form of
such continuum, field theoretic representation of the many-body
hamiltonian is given by a conformal field theory with dynamical
$SU\left(R\right)$ charges \cite{Avan:1996vi}
\begin{eqnarray}
H & = & b\oint z^{2}\left(\frac{1}{6}\left(\alpha\left(z\right)\right)^{3}+\alpha\left(z\right)T^{J}\left(z\right)\right)+c\left(R\right)W\left(z\right) \nonumber\\
 &  & +\oint dz\oint dw\frac{zw}{\left(z-w\right)^{2}}\left(J\left(z\right)\cdot J\left(w\right)+\alpha\left(z\right)\alpha\left(w\right)\right).\label{General form of hamiltonian}\end{eqnarray}
In the above hamiltonian $b$ is a general constant and
$c\left(R\right)$ is a specific $R$-dependent constant.
$\alpha\left(z\right)$ is a bosonic scalar field coupled to a level
$k=1$ current algebra, which participates in the hamiltonian through
a spin-2 $W$-algebra energy-momentum tensor $T^{J}\left(z\right)$
and a spin-3 $W$-algebra generator $W\left(z\right)$, both built
from the $su\left(R\right)$ current algebra \cite{Fukuma:1990yk}.
One should be aware of the fact that the full hamiltonian will have
both $k=1$ and $k=-1$ current algebras, and the corresponding
bosonic $U\left(1\right)$ scalar fields .

For the purposes of this paper we will concern ourselves with the
particular case of $su\left(2\right)$, namely $R=2$, which will be
used to illustrate most of the relevant dynamical features. More details
of the general case of $su\left(R\right)$ and also the large $R$ limit can
be found in
\cite{Avan:1996vi} and \cite{Jevicki:1996wn}. Of considerable future interest is the case
$R=N$ with $N\rightarrow\infty$. Aspects of the
$R=N\rightarrow\infty$ theory were already given in
\cite{Jevicki:1996wn}. Through the large $N$ limit of the current
algebra one realizes a $W_{\infty}$ algebra, the expected symmetry
of the matrix model \cite{Avan:1991ik}.

\section{Current-Algebra Representation}\label{Current-Algebra Rep}

We will now proceed with the study of the Spin Calogero-Sutherland
model by describing in detail its bosonized current algebra
representation. The standard form of the Spin Calogero-Sutherland
model is given by (for example, \cite{ha:1992-}):
\begin{eqnarray} H & = &
-\frac{1}{2}\sum_{i}\frac{\partial^{2}}{\partial
x_{i}^{2}}+\frac{1}{4}\left(\frac{\pi}{L}\right)^{2}\sum_{i\neq
j}\frac{\beta\left(2\beta+1+\vec{\sigma}_{i}\cdot\vec{\sigma}_{j}\right)}{\sin^{2}\left[\frac{\pi}{L}\left(x_{i}-x_{j}\right)\right]}\,,
\label{Original Hamiltonian trigonom}\end{eqnarray} %
where $\frac{2\pi}{L} x_{i}\in\left[0,2\pi\right]$ is the coordinate
on the circle.

 The spin-spin interaction is equivalently written in terms of the
spin exchange operator $P_{ij}$. Recalling that
$J_{i}^{a}=\frac{1}{2}\sigma_{i}^{a}$ for each particle $i$, we have
the relation:\[
P_{ij}=J_{i}^{+}J_{j}^{-}+J_{i}^{-}J_{j}^{+}+2\left(J_{i}^{z}J_{j}^{z}+\frac{1}{4}\right)=\frac{1}{2}\left[\vec{\sigma}_{i}\cdot\vec{\sigma}_{j}+1\right].\]
Consequently, we can rewrite (\ref{Original Hamiltonian trigonom})
as:
\begin{eqnarray} H
& = & \left(\frac{2\pi}{L}\right)^{2}\frac{1}{2}\left\{
\sum_{i}D_{i}^{2}-\sum_{i\neq
j}\frac{z_{i}z_{j}}{\left(z_{i}-z_{j}\right)^{2}}\beta\left(\beta+P_{ij}\right)\right\}
,\label{Original Hamiltonian}\end{eqnarray}%
 where
$D_{i}=z_{i}\frac{\partial}{\partial z_{i}}$, and $z_{i}=e^{i
\frac{2\pi}{L} x_{i}}$. From now on, we will drop the factor of
$\left(\frac{2\pi}{L}\right)$ from the hamiltonian. In general,
models of Calogero-type possess several universal properties. First
of all they describe particles with generalized statistics, governed
by the coupling constant $\beta$, see \cite{Polychronakos:2006nz}.
  Introducing a position exchange
operator $K_{ij}$, we have that $P_{ij}K_{ij}=2x-1$ ($x=+1$ is the
case of boson, while the $x=0$ is for fermions), for the
corresponding wavefunctions \[
\Psi\left(...,z_{i}\,\sigma_{i},...,z_{j}\,\sigma_{j},...\right)=\left(-1\right)^{x+1}\Psi\left(...,z_{j}\,\sigma_{j},...,z_{i}\,\sigma_{i},...\right),\]

For the bosonization method we will have
to consider a system of fermions. In general one can go to a fermionic (or bosonic)
versions of the hamiltonian by applying a similarity transformation\begin{eqnarray*}
\tilde{H} & = & 2\left(\frac{L}{2\pi}\right)^{2}\Delta^{-\beta}H\Delta^{\beta}-E_{0},\end{eqnarray*}
 where \begin{eqnarray*}
\Delta^{\beta} & = & \prod_{i<j}\sin^{\beta}\left(\frac{1}{2}\left(x_{i}-x_{j}\right)\right)=\prod_{i<j}\left(z_{i}-z_{j}\right)^{\beta}\prod_{i}z_{i}^{-\beta\frac{N-1}{2}},\\
E_{0} & = & \beta^{2}\frac{N(N^{2}-1)}{12}.\end{eqnarray*}

 By means of some identities, such as \begin{equation*}
 \frac{z_{i}z_{k}}{\left(z_{j}-z_{i}\right)\left(z_{j}-z_{k}\right)}+\frac{z_{k}z_{j}}{\left(z_{i}-z_{k}\right)\left(z_{i}-z_{j}\right)}+\frac{z_{j}z_{i}}{\left(z_{k}-z_{j}\right)\left(z_{k}-z_{i}\right)}=1\, ,
 \end{equation*}
the similarity transformation given above provides an effective
fermionic hamiltonian of the form \begin{eqnarray} \tilde{H} & = &
\sum_{i}D_{i}^{2}+\frac{\beta}{2}\sum_{i\neq
j}\frac{z_{i}+z_{j}}{z_{i}-z_{j}}\left(D_{i}-D_{j}\right)-\frac{\beta}{2}\sum_{i\ne
j}\frac{z_{i}\,z_{j}}{\left(z_{i}-z_{j}\right)^{2}}\left(3+\vec{\sigma}_{i}\cdot\vec{\sigma}_{j}\right).\label{Awata
effective hamiltonian}\end{eqnarray}

Through standard second quantization, the hamiltonian (\ref{Awata
effective hamiltonian}) is written in terms of fermionic
fields\begin{equation} \widetilde{\Psi}=\left(\begin{array}{c}
\psi_{1}\\
\psi_{2}\end{array}\right)\quad,\qquad J^{i}=\widetilde{\Psi}^{\dagger}\frac{\sigma^{i}}{2}\widetilde{\Psi}\;\left(i=1,2,3\right).\end{equation}
 obeying the anti-commutation relations ($a=1,2$):\[
\left\{ \psi_{a}(x),\psi_{b}(y)\right\} =\delta_{ab}\delta(x-y).\]
Before continuing, we shall review some important aspects of the
bosonization procedure.

 In the abelian bosonization procedure, each fermi field $\psi,\psi^{\dagger}$
is expressed in terms of a boson $\phi$ by the formulas (as
references on bosonization, see
\cite{Fukuma:1990yk,Douglas:1993wy,senechal:1999-}) :\[
\psi\left(\theta\right)=\frac{1}{\sqrt{2\pi}}:e^{-i\phi\left(z\right)}:\qquad,\qquad\psi^{\dagger}\left(\theta\right)=\frac{z}{\sqrt{2\pi}}:e^{i\phi\left(z\right)}:=z\psi^{\dagger}\left(z\right)\,.\]
 As there are two fermion fields $\psi_{a}$, two
bosons $\phi_{a}$ will be needed. Because we have more than one
species of fermionic fields (i.e. two), we also need to employ Klein
factors to ensure the anticommutativity among the species. This will
be particularly relevant when considering the 4-fermion interactions
later.

After normal ordering one has: \begin{eqnarray*}
:\psi^{\dagger}\left(w'\right)\psi\left(w\right): & = & \frac{iw}{2\pi}\partial_{w}\phi,\\
:\psi^{\dagger}\left(w'\right)\partial_{w}\psi\left(w\right): & = & \frac{iw}{2\pi}\left\{ \frac{i}{2}\partial_{w}^{2}\phi+\frac{1}{2}\left(\partial_{w}\phi\right)^{2}\right\} ,\\
:\psi^{\dagger}\left(w'\right)\partial_{w}^{2}\psi\left(w\right): & = & \frac{w}{2\pi}\left\{ \frac{i}{3}\partial_{w}^{3}\phi+\left(\partial_{w}\phi\right)\left(\partial_{w}^{2}\phi\right)-\frac{i}{3}\left(\partial_{w}\phi\right)^{3}\right\} .\end{eqnarray*}

For the $U\left(1\right)$ currents
$\alpha^{a}\left(z\right),\,\left(a=1,2\right)$ associated with each
fermionic field $\psi_{a}\left(z\right)$ we have \begin{equation}
\alpha^{a}\left(z\right)\equiv\psi_{a}^{\dagger}\left(z\right)\psi_{a}\left(z\right)=i\partial_{z}\phi_{a},\end{equation}
 For each of the fermions, the energy-momentum tensor is of the form
\begin{equation}
T^{a}=-\psi_{a}^{\dagger}\partial\psi_{a}=\frac{1}{2}\left(\partial\psi_{a}^{\dagger}\psi_{a}-\psi_{a}^{\dagger}\partial\psi_{a}\right)-\frac{1}{2}\partial\left(\psi_{a}^{\dagger}\psi_{a}\right),\end{equation}
 corresponding to a central charge $c=-2$.%
\footnote{For an energy-momentum tensor of the form $
T=\frac{1}{2}\left(\partial\psi^{\dagger}\psi-\psi^{\dagger}\partial\psi\right)+\mu\partial\left(\psi^{\dagger}\psi\right)$
 the central charge equals $c=1-12\mu^{2}$.%
}

A central role will be played by the $SU\left(2\right)$ currents
given by\begin{eqnarray*}
J^{z} & = & \frac{1}{2}\left[\psi_{1}^{\dagger}\psi_{1}-\psi_{2}^{\dagger}\psi_{2}\right]=\frac{1}{2}J^{1}\left(z\right)-\frac{1}{2}J^{2}\left(z\right)\\
J^{\pm} & = &
\frac{1}{2}\widetilde{\Psi}^{\dagger}\left(\sigma^{1}\pm
i\sigma^{2}\right)\widetilde{\Psi}\quad\Rightarrow\:
J^{+}=\psi_{1}^{\dagger}\psi_{2}\;\,\, , \,\,\;
J^{-}=\psi_{2}^{\dagger}\psi_{1}\,.\end{eqnarray*}
 They obey a level-1 $su(2)$ current algebra with the energy momentum
tensor being \begin{equation}
T\left(z\right)=-\widetilde{\Psi}^{\dagger}\partial\widetilde{\Psi}=-\psi_{1}^{\dagger}\partial\psi_{1}-\psi_{2}^{\dagger}\partial\psi_{2}=T^{1}\left(z\right)+T^{2}\left(z\right).\end{equation}
 This will generate a Virasoro algebra of central charge $c=-4$,
which is expected as we are looking at a theory with two anticommuting
fermions $\psi^{1},\psi^{2}$.

We can proceed with the bosonization of the hamiltonian. We
distinguish three distinct terms in the hamiltonian
as:\begin{equation}
\tilde{H}=\underbrace{{-\sum_{i=1}^{N}\frac{\partial^{2}}{\partial
x_{i}^{2}}}}_{H_{1}}\underbrace{-\frac{\beta}{2}\sum_{i\ne
j}\cot\left(\frac{x_{i}-x_{j}}{2}\right)\left(\frac{\partial}{\partial
x_{i}}-\frac{\partial}{\partial
x_{j}}\right)}_{H_{2}}+\underbrace{\frac{\beta}{4}\sum_{i\ne
j}\sin^{-2}\left(\frac{x_{i}-x_{j}}{2}\right)\left(1+P_{ij}\right)}_{H_{s}}.\label{1st
quantized Uglov}\end{equation}

The first term in the hamiltonian becomes:\begin{eqnarray*}
 H_{1} & \rightarrow & -\int dx\widetilde{\Psi}^{\dagger}\left(x\right)\frac{\partial^{2}}{\partial x^{2}}\widetilde{\Psi}\left(x\right)=\oint\frac{dz}{iz}\widetilde{\Psi}^{\dagger}\left(z\right)\left[z\partial+z^{2}\partial^{2}\right]\widetilde{\Psi}\left(z\right)\\
 & = & \sum_{a}\oint\frac{dz}{2\pi i}\left[z^{2}\left\{
\frac{i}{3}\partial^{3}\phi_{a}+\partial\phi_{a}\partial^{2}\phi_{a}-\frac{i}{3}\left(\partial\phi_{a}\right)^{3}\right\}
+z\left\{
\frac{i}{2}\partial^{2}\phi_{a}+\frac{1}{2}\left(\partial\phi_{a}\right)^{2}\right\}
\right]\,.\end{eqnarray*}

In the case of the second term, we have to be more careful since one
has a pole at coincident points (this we handle by point-splitting
regularization): \begin{eqnarray*}  H_{2} & \rightarrow & -\beta\int
dx\int
dy\widetilde{\Psi}^{\dagger}\left(x\right)\widetilde{\Psi}^{\dagger}\left(y\right)\cot\left(\frac{x-y}{2}\right)\widetilde{\Psi}\left(y\right)\frac{\partial}{\partial
x}\widetilde{\Psi}\left(x\right)+\beta\int
dx\widetilde{\Psi}^{\dagger}\left(x\right)\frac{\partial^{2}}{\partial
x^{2}}\widetilde{\Psi}\left(x\right).\end{eqnarray*}
 Note that the last term is just $H_{1}$, bosonized before. Going
to complex coordinates,\begin{eqnarray*}
 H_{2} & = & -\beta\sum_{a,b}\oint\frac{dz}{z}\oint\frac{dw}{w}\frac{z+w}{z-w}\psi_{a}^{\dagger}\left(z\right)\psi_{b}^{\dagger}\left(w\right)\psi_{b}\left(w\right)z\partial_{z}\psi_{a}\left(z\right)-\beta H_{1}\\
  & = & \beta\sum_{a,b}\oint\frac{dz}{2\pi i}\oint\frac{dw}{2\pi i}\left(\frac{z+w}{z-w}\right)i\partial_{w}\phi_{b}z\left[\frac{i}{2}\partial_{z}^{2}\phi_{a}+\frac{1}{2}\left(\partial_{z}\phi_{a}\right)^{2}\right]-\beta H_{1},\end{eqnarray*}
 where the contour integral of $w$ has to be averaged over an outer
and an inner circles around the $z$ contour.

Finally we come to the bosonization of the last term $H_{s}$ of our
Hamiltonian:\begin{eqnarray*}
 H_{s} & = & \frac{\beta}{8}\sum_{i\ne j}\sin^{-2}\left(\frac{x_{i}-x_{j}}{2}\right)\left(3+\vec{\sigma}_{i}\cdot\vec{\sigma}_{j}\right)\\
  & \rightarrow & \frac{3\beta}{8}\int dx\int dy\widetilde{\Psi}^{\dagger}\left(x\right)\widetilde{\Psi}^{\dagger}\left(y\right)\sin^{-2}\left(\frac{x-y}{2}\right)\widetilde{\Psi}\left(y\right)\widetilde{\Psi}\left(x\right)+\\
  &  & \qquad\qquad\qquad+\frac{\beta}{8}\sum_{k=1}^{3}\int dx\int dy\sin^{-2}\left(\frac{x-y}{2}\right)\widetilde{\Psi}^{\dagger}\left(x\right)\widetilde{\Psi}^{\dagger}\left(y\right)\sigma^{k}\widetilde{\Psi}\left(y\right)\sigma^{k}\widetilde{\Psi}\left(x\right).\end{eqnarray*}
 Again we encounter a singularity, the pole is now of second order,
and so a regularization gives a zero contribution. Going to the
complex plane, \begin{eqnarray*}  H_{s} & = &
-\frac{3\beta}{2}\sum_{a,b}\oint\frac{dz}{2\pi i}\oint\frac{dw}{2\pi
i}\frac{i\partial_{z}\phi_{a}\, zw\,
i\partial_{w}\phi_{b}}{\left(z-w\right)^{2}}+\frac{\beta}{2}\sum_{k}\int
dx\int
dy\frac{\widetilde{\Psi}^{\dagger}\left(y\right)\frac{\sigma^{k}}{2}\widetilde{\Psi}\left(y\right)\widetilde{\Psi}^{\dagger}\left(x\right)\frac{\sigma^{k}}{2}\widetilde{\Psi}\left(x\right)}{\sin^{2}\left(\frac{x-y}{2}\right)},\end{eqnarray*}
where the integration on $w$ is to be done around the $z$ contour.
Important to point out is the bosonized representation for the spin
permutation operator $P_{ij}$:\begin{eqnarray*}
 -\beta\sum_{i\ne j}\frac{z_{i}z_{j}}{\left(z_{i}-z_{j}\right)^{2}}P_{ij} & \rightarrow & \beta\sum_{a,b}\oint dz\oint dw\frac{\psi_{b}^{\dagger}(z)\psi_{a}^{\dagger}(w)\psi_{b}(w)\psi_{a}(z)}{(z-w)^{2}}\\
  & = & \beta\sum_{a,b}\oint\frac{dz}{2\pi i}\oint\frac{dw}{2\pi i}:e^{-i(\phi_{a}(w)-\phi_{b}(w))}:\frac{zw}{(z-w)^{2}}:e^{-i(\phi_{b}(z)-\phi_{a}(z))}:\, ,\end{eqnarray*}
because it has a very similar form to the one that will be found in
the next section by collective field theory methods.

Returning to the full hamiltonian, its final form in the bosonized
representation reads:\begin{eqnarray*}
 \tilde{H} & = & \frac{1-\beta}{6}\oint\frac{dz}{2\pi i}\left[2z^{2}\alpha_{a}^{3}+3z\alpha_{a}^{2}+\alpha_{a}\right]+\frac{\beta}{2}\oint\frac{dz}{2\pi i}\oint\frac{dw}{2\pi i}\,\frac{z+w}{z-w}\alpha_{b}\left(w\right)z\left[\partial\alpha_{a}\left(z\right)-z\alpha_{a}^{2}\left(z\right)\right]\\
  &  & -\frac{3\beta}{2}\oint\frac{dz}{2\pi i}\oint\frac{dw}{2\pi i}\frac{zw}{\left(z-w\right)^{2}}\alpha_{a}\left(z\right)\alpha_{b}\left(w\right)+2\beta\oint dz\oint dw\frac{zw}{\left(z-w\right)^{2}}\sum_{k=1}^{3}J^{k}\left(w\right)J^{k}\left(z\right)\,.\end{eqnarray*}
 The indices $a,b$ are summed over, and $\alpha_{a}=i\partial\phi_{a}$. We can also express the
Hamiltonian in terms of the spin $\beta$ and charge $B$
bosons,\footnote{We trust the reader will be able to distinguish
between the statistics parameter and the spin field, both labeled
$\beta$. The latter will be supplemented with an argument, for
example $\beta\left(z\right)$, in order to avoid confusion.} given
by:\begin{equation}
B\left(x\right)=\alpha_{1}\left(x\right)+\alpha_{2}\left(x\right)\qquad;\qquad\beta\left(x\right)=\alpha_{1}\left(x\right)-\alpha_{2}\left(x\right).\end{equation}
 With these we have: \begin{eqnarray}
 \tilde{H} & = & \frac{1-\beta}{12}\oint\frac{dz}{2\pi i}\left[z^{2}B\left(z\right)\left(B^{2}\left(z\right)+3\beta^{2}\left(z\right)\right)+3z\left(B^{2}\left(z\right)+\beta^{2}\left(z\right)\right)+2B\left(z\right)\right]-\nonumber \\
  &  & +\frac{\beta}{2}\oint\frac{dz}{2\pi i}\oint\frac{dw}{2\pi i}\left(\frac{z+w}{z-w}\right)B\left(w\right)\left[z\partial B\left(z\right)-\frac{1}{2}z\left(B^{2}\left(z\right)+\beta^{2}\left(z\right)\right)\right]+\label{Bosonized Hamiltonian}\\
  &  & -\frac{3\beta}{2}\oint\frac{dz}{2\pi i}\oint\frac{dw}{2\pi i}\frac{zw}{\left(z-w\right)^{2}}B\left(z\right)B\left(w\right)+2\beta\oint dz\oint dw\frac{zw}{\left(z-w\right)^{2}}\sum_{k=1}^{3}J^{k}\left(w\right)J^{k}\left(z\right)\,.\nonumber \end{eqnarray}

We can see in this final expression the general structure given in
equation (\ref{General form of hamiltonian}). We will discuss in
more detail a comparison of various representation further below. We
will furthermore see that the same form of the hamiltonian can be
obtained by collective field techniques as a second way of
determining the bosonized hamiltonian. Finally, the usefulness of
the bosonized representation
 will be
demonstrated in selected applications involving studies of nontrivial
classical background.

\subsection{Mode Expansion}

It is useful to present also the mode expansion of the bosonized Hamiltonian.
We start by introducing the mode expansions for the fermionic fields
$\psi_{a}$ ($a=1,2$):\begin{equation}
\psi_{a}^{\dagger}\left(z\right)=\sum_{n}\psi_{-n}^{a\dagger}z^{-n-1}\quad,\quad\psi_{a}\left(z\right)=\sum_{n}\psi_{n}^{a}z^{-n},\label{5.2 Expansion of psi}\end{equation}
 with commutation relations $\left\{ \psi_{n}^{a},\psi_{m}^{b\dagger}\right\} =\delta_{m,n}\delta^{ab}\,.$
We also need the expansion of the $U\left(1\right)$ currents:
\begin{equation}
\alpha_{a}\left(z\right)\equiv\psi_{a}^{\dagger}\left(z\right)\psi_{a}\left(z\right)=i\partial_{z}\phi_{a}=\sum_{n}\alpha_{n}^{a}z^{-n-1},\label{5.2
Current for fermions}\end{equation}
  with commutation relations
$\left[\alpha_{n}^{a},\alpha_{m}^{b}\right]=n\delta^{ab}\delta_{m+n,0}$.
The components of the current $\alpha^{a}$ can be written in terms
of the fermionic field modes as:\begin{equation}
\alpha_{n}^{a}=\oint\frac{dz}{2\pi
i}z^{n}\psi_{a}^{\dagger}\left(z\right)\psi_{a}\left(z\right)=\sum_{\ell,m}\oint\frac{dz}{2\pi
i}z^{n-\left(\ell+m\right)-1}\psi_{-\ell}^{a\dagger}\psi_{m}^{a}=\sum_{m}\psi_{m-n}^{a\dagger}\psi_{m}^{a}\,.\label{5.2
Expansion of current in psi modes}\end{equation}

 For each of the fermions we have the energy-momentum tensor, \begin{equation}
T^{a}=-\psi_{a}^{\dagger}\partial\psi_{a}=\sum_{n}L_{n}^{a}z^{-n-2}.\label{5.2 EM tensor for fermions -2 charge}\end{equation}
 The Virasoro generators $L_{n}^{a}$ are given by:\begin{equation}
 L_{n}^{a}=-\frac{1}{2}\oint\frac{dz}{2\pi
i}z^{n+1}\psi_{a}^{\dagger}\partial\psi_{a}=-\sum_{\ell,m}\oint\frac{dz}{2\pi
i}z^{n-\left(\ell+m\right)-1}\left(-m\right)\psi_{-\ell}^{a\dagger}\psi_{m}^{a}=\sum_{m}m\psi_{m-n}^{a\dagger}\psi_{m}^{a}.\label{5.2
Virasoro fermions -2 charge}\end{equation}

 Consider next  the operators of a $su\left(2\right)$ current algebra
 $J^{i}\left(z\right)$ ($i,j,k=1,2,3\equiv x,y,z$). These can be expanded as
$J^{i}\left(z\right)=\frac{1}{2\pi}\sum_{N}J_{N}^{i}z^{-N-1}$,
with\begin{eqnarray}
2\pi J_{n}^{x} & = & \frac{1}{2}\left(J^{+}+J^{-}\right)_{n}=\frac{1}{2}\sum_{m}\left[\psi_{\left(m-n\right)}^{1\dagger}\psi_{m}^{2}+\psi_{\left(m-n\right)}^{2\dagger}\psi_{m}^{1}\right]\nonumber \\
2\pi J_{n}^{y} & = & \frac{1}{2i}\left(J^{+}-J^{-}\right)_{n}=-\frac{i}{2}\sum_{m}\left[\psi_{\left(m-n\right)}^{1\dagger}\psi_{m}^{2}-\psi_{\left(m-n\right)}^{2\dagger}\psi_{m}^{1}\right]\nonumber \\
2\pi J_{n}^{z} & = &
\frac{1}{2}\left(\alpha_{n}^{1}-\alpha_{n}^{2}\right)=-\frac{1}{2}\sum_{m}\left[\Psi_{\left(m-n\right)}^{2\dagger}\psi_{m}^{2}-\psi_{\left(m-n\right)}^{1\dagger}\psi_{m}^{1}\right]\label{5.2
SU(2) currents}\end{eqnarray} with the commutation
relations\begin{equation}
\left[J_{n}^{i},J_{m}^{j}\right]=i\varepsilon^{ijk}J_{m+n}^{k}+\frac{n}{2}\delta^{ij}\delta_{m+n,0}\,.\label{5.2
SU(2) current [ , ]}\end{equation}

 We already introduced the
\emph{charge field} (the $B$ field) and the \emph{spin field} (the
$\beta$ field), and saw that to bosonize the hamiltonian, we needed
two scalar bosonic fields $\alpha_{1},\alpha_{2}$, because we were
interested in the $SU\left(2\right)$ current algebra. If we now
identify $\alpha_{i}\equiv\alpha^{\sigma_{i}}$ , where $\sigma=\pm1$
(or alternatively $\sigma_{1}=\uparrow$ and
$\sigma_{2}=\downarrow$), then the following expansions hold, in
terms of $z$ coordinates:
\begin{equation} \left\{ \begin{array}{l}
B\left(z\right)=\sum_{n}B_{n}z^{-n-1}\\
\\\beta\left(z\right)=\sum_{n}\beta_{n}z^{-n-1}\end{array}\right.,\;\rm{where}\quad\left\{ \begin{array}{l}
B_{n}=\sum_{\sigma}\alpha_{n}^{\sigma}\\
\\\beta_{n}=\sum_{\sigma}\sigma\alpha_{n}^{\sigma}\end{array}\right..\label{Def: charge and spin fields}\end{equation}

In terms of $su\left(2\right)$ current algebra, given the two
fermionic fields $\psi_{1}=\psi_{\uparrow}$ and
$\psi_{2}=\psi_{\downarrow}$, the spinor field
$\Psi=\left(\psi_{1}\:\psi_{2}\right)^{T}$ and the currents
$J^{k}=\Psi^{\dagger}\frac{\sigma^{k}}{2}\Psi$, the total theory has
an energy momentum tensor with central charge $c=-4$. Its
decomposition into a level $k=1$ $su\left(2\right)$ current algebra
$\left\{ J_{m}^{k}\right\} $ and a Heisenberg algebra $\left\{
B_{n}\right\} $ is such that $T_{m}$ is composed of two Virasoro
algebras, one related to the currents $J$, with charge $c=1$, and
one related to {}``free'' bosons with charge $c=-5$\[
T_{m}=L_{m}^{J}+L_{m}^{B},\]
 where

\begin{eqnarray}
L_{m}^{B} & = & \frac{1}{4}\sum_{n\in\mathbb{Z}}:B_{-n}B_{n+m}:+\frac{1}{2}\left(m+1\right)B_{m}\label{Virasoro generators for free bosons}\\
L_{m}^{J} & = & \frac{1}{4}\sum_{e\in\mathbb{Z}}:\beta_{-m}\beta_{n+m}:\:.\label{Virasoro generators for currents}\end{eqnarray}

We can now expand the bosonized hamiltonian in terms of modes. We go
back to the expression for the bosonized hamiltonian (\ref{Bosonized
Hamiltonian}), and, by separating the several terms and by making
use of the expansion (\ref{Def: charge and spin fields}), the first
term becomes:\begin{eqnarray*}  H_{1} & \equiv &
\left(1-\beta\right)\left\{
\left(B_{0}+1\right)\left(\frac{1}{3}L_{0}^{B}+L_{0}^{J}\right)+\sum_{n\geq1}\left(B_{-n}\left(\frac{1}{3}L_{n}^{B}+L_{n}^{J}\right)+\left(\frac{1}{3}L_{-n}^{B}+L_{-n}^{J}\right)B_{n}\right)\right\}
.\end{eqnarray*}
 For the second term we have:\begin{eqnarray*}
 H_{2} & \equiv & \frac{\beta}{2}\left\{
2\sum_{n\geq1}nB_{-n}B_{n}-2\sum_{k\geq1}\left(L_{-k}^{B}B_{k}-B_{-k}L_{k}^{B}+kB_{-k}B_{k}\right)-2\sum_{k\geq1}\left(L_{-k}^{J}B_{k}-B_{-k}L_{k}^{J}\right)\right\}
.\end{eqnarray*}
 Finally, the third term will become\begin{eqnarray*}
 H_{3} & \equiv &
-\frac{3\beta}{2}\sum_{n\geq1}nB_{-n}B_{n}-2\beta\sum_{k=1}^{3}\sum_{n\geq1}nJ_{-n}^{k}J_{n}^{k}\,.\end{eqnarray*}

The final result for the total hamiltonian in terms of boson (and
current) modes is:\begin{eqnarray}
 \tilde{H} & = & \left(\beta-1\right)\left(B_{0}+1\right)\left(\frac{1}{3}L_{0}^{B}+L_{0}^{J}\right)+\frac{1+2\beta}{3}\sum_{n\geq1}B_{-n}L_{n}^{B}+\frac{1-4\beta}{3}\sum_{n\geq1}L_{-n}^{B}B_{n}+\nonumber \\
  &  & \qquad+\sum_{n\geq1}B_{-n}L_{n}^{J}+\left(1-2\beta\right)\sum_{n\geq1}L_{-n}^{J}B_{n}-\frac{3\beta}{2}\sum_{n\geq1}nB_{-n}B_{n}-2\beta\sum_{n\geq1}nJ_{-n}\cdot J_{n}.\label{Hamiltonian in terms of modes}\end{eqnarray}

We now comment on the relation of the above bosonized Hamiltonian
to the various bosonized versions of the model studied in the literature.
First, by a simple change of our notation :\begin{equation}
\beta\rightarrow-\frac{1}{\alpha}\quad;\qquad H_{U}=\alpha\tilde{H}.\label{Comparison with bosonization from Uglov}\end{equation}
 we obtain a form analogous to one constructed by Uglov in \cite{Uglov:1996ba}\begin{eqnarray}
 H_{U} & = & E_{0}+\left(\alpha+1\right)\left(B_{0}+1\right)\left(\frac{1}{3}L_{0}^{B}+L_{0}^{J}\right)+\frac{\alpha-2}{3}\sum_{n\geq1}B_{-n}L_{n}^{B}+\frac{\alpha+4}{3}\sum_{n\geq1}L_{-n}^{B}B_{n}\nonumber \\
  &  & +\alpha\sum_{n\geq1}B_{-n}L_{n}^{J}+\left(\alpha+2\right)\sum_{n\geq1}L_{-n}^{J}B_{n}+\frac{3}{2}\sum_{n\geq1}nB_{-n}B_{n}+2\sum_{k}\sum_{n\geq1}nJ_{-n}^{k}J_{n}^{k}\,,\label{3 - Uglov Hamiltonian}\end{eqnarray}
with\begin{equation}
\left[B_{n},B_{m}\right]=2n\delta_{n+m,0}\,.\end{equation}

One can also see that the bosonized hamiltonian  described above
fits into  a (more symmetric) form of the representation originally
constructed in \cite{Avan:1996vi} which reads:

\begin{equation}
H_{AJL}=\frac{b}{6q}\sum_{n_{i}}\alpha_{n_{1}}\alpha_{n_{2}}\alpha_{n_{3}}\delta_{n_{1}+n_{2}+n_{3},0}+b\sum_{n}\alpha_{n}L_{-n}^{J}+\frac{1}{q}\sum_{n>0}n\alpha_{-n}\alpha_{n}+\sum_{n>0}nJ_{-n}^{k}J_{n}^{k}.\label{Hamiltonian
Avan+Jevicki}\end{equation}
This is the most general form of the hamiltonian, and the modes $\alpha_{n}$ obey
the commutation relation:\begin{equation}
\left[\alpha_{n},\alpha_{m}\right]=qn\delta_{m+n,0},\end{equation}

In order to exhibit  the precise comparison, we will collect terms
which have the same number of creation-annihilation operators:
 starting from (\ref{3 - Uglov Hamiltonian}), we denote the cubic
term of $H_{U}$ by $H_{U}^{3}$, the quadratic term in $B_{n}$ by
$H_{U}^{2}$, the terms linear in $B_{n}$ and $L_{n}^{J}$ by $
H_{U}^{1,J}$, and finally the pure current term will be denoted by
$H_{U}^{J}$.

 The following change in the normalization of the fields and  their
commutation relations will be needed perform the
comparison:\begin{eqnarray*}
B'_{n} & = & \left(\alpha+2\right)B_{n},\qquad n\geq1\\
B'_{-n} & = & \alpha B_{-n}\\
\left[B'_{n},B'_{m}\right] & = & 2\alpha(\alpha+2)n\delta_{n+m,0}\,.\end{eqnarray*}
 From now on, we are going to drop the primes and will use the same
notation for the new fields. In terms of the new fields, $H_{U}$
is given by the sum of the following terms:\begin{eqnarray}
H_{U}^{3} & = & \frac{1}{4\alpha\left(\alpha+2\right)}\sum_{n\geq1}\left(B_{-n}B_{-m}B_{n+m}+B_{-n-m}B_{n}B_{m}\right)\,,\label{3 - Cubic Uglov}\\
H_{U}^{2} & = & \frac{1}{2\alpha\left(\alpha+2\right)}\sum_{n\geq1}nB_{-n}B_{n}+\frac{\alpha+1}{2\alpha(\alpha+2)}\left(B_{0}+1\right)\sum_{n\geq1}B_{-n}B_{n}\,,\label{3 - Quadratic Uglov}\\
H_{U}^{1,J} & = & \sum_{n\geq1}\left(B_{-n}L_{n}^{J}+L_{-n}^{J}B_{n}\right)+\left(\alpha+1\right)\left(B_{0}+1\right)L_{0}^{J}\,,\label{3 - Linear Uglov}\\
H_{U}^{J} & = & 2\sum_{n\geq1}\sum_{k}nJ_{-n}^{k}J_{n}^{k}\,.\label{3 - Current Uglov}\end{eqnarray}

Now, let us consider the hamiltonian (\ref{Hamiltonian
Avan+Jevicki}). We will use a similar notation to refer to the
cubic, quadratic, linear and pure current terms in the hamiltonian.
Changing the normalization of the $\alpha_{n}$\begin{eqnarray*}
\alpha'_{n} & = & b\alpha_{n},\qquad\forall n\\
\left[\alpha'_{n},\alpha'_{m}\right] & = & b^{2}qn\delta_{n+m,0}\,,\end{eqnarray*}
 and again dropping the primes on the new fields, $H_{AJL}$ will then be the
sum of:

\begin{eqnarray}
H_{AJL}^{3} & = & \frac{1}{2b^{2}q}\sum_{n,m\geq1}\left(\alpha_{-n}\alpha_{-m}\alpha_{n+m}+\alpha_{-n-m}\alpha_{n}\alpha_{m}\right)\,,\label{3 - Cubic AJL}\\
H_{AJL}^{2} & = & \frac{1}{b^{2}q}\sum_{n\geq1}n\alpha_{-n}\alpha_{n}+\frac{1}{b^{2}q}\alpha_{0}\sum_{n\geq1}\alpha_{-n}\alpha_{n}\,,\label{3 - Quadratic AJL}\\
H_{AJL}^{1,J} & = & \sum_{n\geq1}\left(\alpha_{-n}L_{n}^{J}+L_{-n}^{J}\alpha_{n}\right)+\alpha_{0}L_{0}^{J}\,,\label{3 - Linear AJL}\\
H_{AJL}^{J} & = & \sum_{n\geq1}\sum_{k}nJ_{-n}^{k}J_{n}^{k}\,.\label{3 - Current AJL}\end{eqnarray}
 The relation between $B_{n}$ and $\alpha_{n}$ can be determined.
In fact, if the following relation holds, the two hamiltonians
become equivalent with the same commutation
relations.\begin{eqnarray*}
b^{2}q & = & 2\alpha(\alpha+2)\\
\alpha{}_{0} & = & \left(\alpha+1\right)\left(B_{0}+1\right)\,.\end{eqnarray*}
 The factor of $2$ in $H_{U}^{J}$ is not present in $H_{AJL}^{J}$.
That is due to a different definition of the current modes. We can
see that there is perfect agreement between the two hamiltonians.
With this we have also demonstrated that the bosonized hamiltonian  $H_{U}$,  is hermitian.

One last comment should be made. If we write the hamiltonian
(\ref{Hamiltonian in terms of modes}) in terms of the original
bosonization fields $\alpha^{\sigma}\equiv\phi^{\sigma}$ we will be
able to make contact with the collective field approach which will
be summarized in section \ref{Collective FT}. As a final comment on
the current algebra representation we mention the following. Here we
have described in detail the formalism in the simplest SU(2) case
for purposes of comparing different approaches and describing some
nontrivial applications. As we have mentioned the formalism for
general SU(R) was developed in \cite{Avan:1996vi} . Of considerable
interest is the case $R=N$ with $N\rightarrow\infty$. The current
algebraic representation is well suited for that since it leads to a
large N limit of WZW models, a well studied subject. Through the
large $N$ limit of the current algebra one realizes a $W_{\infty}$
algebra, the expected symmetry of the matrix model
\cite{Avan:1991ik}.Aspects of the $R=N\rightarrow\infty$ theory were
already given in \cite{Jevicki:1996wn}
 .

\section{Collective Field Theory}\label{Collective FT}

 For obtaining the collective field theory representation one considers a
permutation symmetric version of the model with the idea of
expressing the theory in terms of collective densities. Considering the
original hamiltonian (\ref{Original Hamiltonian}), given by:\[
H=\sum_{i}D_{i}^{2}-\sum_{i\neq
j}\frac{z_{i}z_{j}}{\left(z_{i}-z_{j}\right)^{2}}\beta\left(\beta+P_{ij}\right),\]
we therefore  need to perform a similarity transformation to bring
the theory  into a bosonic picture. We write the wavefunctions of $H$ as
 $\Psi=\Psi_{0}\chi$, with $\chi$ being a bosonic function. The form of
$\Psi_{0}$ defines then the statistics of the model. In the previous
construction we performed a transformation to the fermionic picture
by using $\Psi_{0}=\Delta^{\beta}$. To obtain a bosonic picture, we
will now require an extra determinant factor in the similarity
transformation, $\Psi_{0}=\Delta^{\beta+1}$. In that case, the
gauged hamiltonian is:\begin{equation}
H_{\beta+1}=\underbrace{\sum_{i}D_{i}^{2}}_{H_{\beta+1}^{1}}+\underbrace{\frac{\left(\beta+1\right)}{2}\sum_{i\ne
j}\frac{z_{i}+z_{j}}{z_{i}-z_{j}}\left(D_{i}-D_{j}\right)}_{H_{\beta+1}^{2}}\underbrace{-2\beta\sum_{i<j}\theta_{ij}\theta_{ji}\left[1-P_{ij}\right]}_{H_{\beta+1}^{s}}+E_{GS}\left(\beta+1\right),\label{Hamiltonian
beta+1}\end{equation}
 where $\theta_{ij}\equiv\frac{z_{i}}{z_{i}-z_{j}}$. The ground state energy is given by
 $E_{GS}\left(\beta+1\right)=\left(\beta+1\right)^{2}\frac{N\left(N^{2}-1\right)}{12}$.

In the last section we bosonized the hamiltonian $\tilde{H}$ in
terms of current fields \cite{Avan:1995sp,Avan:1996vi}. In this
section we describe in some detail the collective field theory
approach. This method was useful for various matrix model problems
and in particular systems of Calogero-type
\cite{Andric:1982jk,Feinberg:2004vz,Bardek:2004yj}. In the case of a
spin-Calogero model one can similarly obtain the hamiltonian in
terms of collective modes in momentum space
\cite{Awata:1995by}.\footnote{Note that the gauged hamiltonian found
in \cite{Awata:1995by} is related to (\ref{Hamiltonian beta+1}) by
changing, in the latter, $\beta+1$ to $\beta$ on the second term.
The discrepancy comes from a difference on the starting point of each case.} %
We will also want to obtain the same result in terms of collective
fields in coordinate space, as it will be useful in further
perturbative and nonperturbative
 studies.


We will start from the hamiltonian $H_{\beta+1}$ (one could get a
similar result by starting from $\tilde{H}\equiv H_{\beta}$). One
defines collective densities
\begin{equation}
\phi_{\sigma}\left(x\right)=\sum_{i=1}^{N}\delta\left(x-x_{i}\right)\delta_{\sigma\sigma_{i}}\quad,\quad\phi\left(x\right)=\sum_{\sigma}\phi_{\sigma}\left(x\right)\end{equation}
 and its conjugate momentum
 $\Pi^{\sigma}\left(x\right)=\frac{\delta}{\delta\phi_{\sigma}\left(x\right)}$.\footnote{In this section coordinate $x$ has range on the interval
$\left[-\pi,\pi\right]$ unless said otherwise. Also, the spin
variable $\sigma$ has possible values $\sigma=\pm 1$.}%
 We take $\Pi^{\sigma}\left(x\right)$ to be non-hermitian just to
save writing $i$s. In terms of Fourier modes, the fields defined
above become (with $n$ integer because we are on the
 circle):
\begin{eqnarray*}
 \phi_{n}^{\sigma}=\int_{-\pi}^{\pi}dx
e^{inx}\phi_{\sigma}\left(x\right)=\sum_{i=1}^{N}z_{i}^{n}\delta_{\sigma\sigma_{i}}\quad
& , & \qquad
\Pi_{n}^{\sigma}=\frac{\partial}{\partial\phi_{n}^{\sigma}}=\int\frac{dx}{2\pi}e^{-inx}\frac{\partial}{\partial\phi_{\sigma}\left(x\right)},\end{eqnarray*}

The expressions that follow are obtained through a change of
variables following the standard collective field techniques
\cite{Jevicki:1979mb}. We start from (\ref{Awata effective
hamiltonian}) and define $\hat{\Pi}_{n}^{\sigma}\equiv
n\frac{\partial}{\partial\phi_{n}^{\sigma}}=n\Pi_{n}^{\sigma}$ (such
that
$\left[\phi_{n}^{\sigma},\hat{\Pi}_{m}^{\sigma'}\right]=-m\delta_{\sigma\sigma'}\delta_{m,n}$).
Writing the dependence on spin explicitly, the derivative $D_{i}$ is
given by chain rule:\[  D_{i}=z_{i}\frac{\partial}{\partial
z_{i}}=\sum_{\sigma}\sum_{n}z_{i}\frac{\partial\phi_{n}^{\sigma}}{\partial
z_{i}}\frac{\partial}{\partial\phi_{n}^{\sigma}}=\sum_{\sigma}\sum_{n}z_{i}\frac{\partial}{\partial
z_{i}}\left(\sum_{j}z_{j}^{n}\delta_{\sigma\sigma_{j}}\right)\frac{\partial}{\partial\phi_{n}^{\sigma}}=\sum_{\sigma}\sum_{n}z_{i}^{n}\delta_{\sigma\sigma_{i}}\hat{\Pi}_{n}^{\sigma}.\]
 This implies that for $H_{1}$ we have
 \begin{eqnarray*}
 H_{1} & = & \sum_{i}D_{i}^{2}=\sum_{\sigma}\sum_{i}\sum_{n}\delta_{\sigma\sigma_{i}}z_{i}\left\{ \frac{\partial z_{i}^{n}}{\partial z_{i}}\hat{\Pi}_{n}^{\sigma}+z_{i}^{n}\sum_{\sigma'}\sum_{m}\frac{\partial\phi_{m}^{\sigma'}}{\partial z_{i}}\frac{\partial}{\partial\phi_{m}^{\sigma'}}\hat{\Pi}_{n}^{\sigma}\right\} \\
  & = & \sum_{\sigma}\left\{ \sum_{n}n\phi_{n}^{\sigma}\hat{\Pi}_{n}^{\sigma}+\sum_{m,n}\phi_{n+m}^{\sigma}\hat{\Pi}_{m}^{\sigma}\hat{\Pi}_{n}^{\sigma}\right\} .\end{eqnarray*}
 Similarly for $H_{2}$
 \begin{eqnarray*}
 H_{2} & = & \frac{\beta+1}{2}\sum_{i\ne j}\frac{z_{i}+z_{j}}{z_{i}-z_{j}}\left(D_{i}-D_{j}\right)\\
  & = & \left(\beta+1\right)\left(\sum_{\sigma=\sigma'}+\sum_{\sigma\ne\sigma'}\right)\sum_{i\ne,j}\frac{z_{i}}{z_{i}-z_{j}}\sum_{m}\left(z_{i}^{m}\delta_{\sigma\sigma_{i}}\delta_{\sigma'\sigma_{j}}-z_{j}^{m}\delta_{\sigma'\sigma_{i}}\delta_{\sigma\sigma_{j}}\right)\hat{\Pi}_{m}^{\sigma}\,.\end{eqnarray*}
 The term with $\sigma=\sigma'$ is:\begin{eqnarray*}
 H_{\sigma}^{2} & = & \left(\beta+1\right)\sum_{\sigma}\left\{
\sum_{n,m\geq1}\phi_{n}^{\sigma}\phi_{m}^{\sigma}\hat{\Pi}_{m+n}^{\sigma}+\sum_{m\geq1}\left(\phi_{0}^{\sigma}-m\right)\phi_{m}^{\sigma}\hat{\Pi}_{m}^{\sigma}\right\}
.\end{eqnarray*}
 But the term with $\sigma\ne\sigma'$ is more complicated. Doing some algebra leads to (notice that $\phi_{n}$ is
not defined for $n$ negative):\begin{eqnarray*}
H_{\sigma\ne\sigma'}^{2} & = &
\left(\beta+1\right)\sum_{\sigma\ne\sigma'}\left(\sum_{m,n\geq1}\phi_{n}^{\sigma'}\phi_{m}^{\sigma}\hat{\Pi}_{m+n}^{\sigma}+\sum_{m>0}\phi_{0}^{\sigma}\phi_{m}^{\sigma'}\hat{\Pi}_{m}^{\sigma}\right)\,.\end{eqnarray*}

A related result can be found in \cite{Awata:1995by}. In fact the
terms written above agree with the corresponding ones from
\cite{Awata:1995by}.

One is also interested in the expressions for the hamiltonian in
coordinate space, as these will play an important role when
performing small fluctuations around a classical background. This
will be explored further in the sections below. For now let us
express the hamiltonian in terms of density fields.
For the $H_{1}$ we get%
\begin{eqnarray}
 H_{1} & = & -\sum_{i}\frac{\partial^{2}}{\partial x_{i}^{2}}=
-\sum_{\sigma}\int
dx\,\phi_{\sigma}\left(x\right)\partial^{2}\Pi^{\sigma}\left(x\right)-\sum_{\sigma}\int
dx\phi_{\sigma}\left(x\right)\partial\Pi^{\sigma}\left(x\right)\partial\Pi^{\sigma}\left(x\right).\label{Define
H1}\end{eqnarray}
 To determine $H_{2}$ we need to write it as \begin{eqnarray}
 H_{2} & = &
\underbrace{\frac{\beta+1}{2}\sum_{i,j}\left(\theta_{ij}-\theta_{ji}\right)\left(D_{i}-D_{j}\right)}_{H_{21}}\underbrace{-\frac{\beta+1}{2}\sum_{i=j}\lim_{z_{i}\rightarrow
z_{j}}\frac{z_{i}+z_{j}}{z_{i}-z_{j}}\left(D_{i}-D_{j}\right)}_{H_{22}},\label{Define
H21 H22}\end{eqnarray}
 and each term can be found to be:\begin{eqnarray*}
 H_{21} & = & -\frac{\beta+1}{2}\sum_{\sigma,\sigma'}\int dx\phi_{\sigma}\left(x\right)\left(\int dy\phi_{\sigma'}\left(y\right)\cot\left(\frac{x-y}{2}\right)\right)\left[\partial\Pi^{\sigma}\left(x\right)-\partial\Pi^{\sigma'}\left(y\right)\right],\\
 H_{22} & = & \left(\beta+1\right)\int
dx\left(\partial^{2}\phi_{\sigma}\right)\left(x\right)\Pi^{\sigma}\left(x\right)=\left(\beta+1\right)\int
dx\phi_{\sigma}\left(x\right)\partial^{2}\Pi^{\sigma}\left(x\right).\end{eqnarray*}

The term $H_{s}$ can also be expressed in terms of collective
fields. The operator $P_{ij}^{\sigma\sigma'}$ can be defined by how
it acts on $\phi_{n}^{\sigma}$. We know that the effective
hamiltonian acts on wavefunctions that depend solely on
$\phi_{n}^{\sigma}$ (bosonic functions), so we want to determine the
action of $P_{ij}$ on $\chi\left(\phi_{n}^{\sigma}\right)$.
First,\begin{eqnarray*} P_{ij}\phi_{n}^{\sigma} & = &
P_{ij}\sum_{\ell=1}^{N}z_{\ell}^{n}\delta_{\sigma\sigma_{\ell}}
 =\phi_{n}^{\sigma}+z_{i}^{n}\left(\delta_{\sigma,\sigma_{j}}-\delta_{\sigma,\sigma_{i}}\right)+z_{j}^{n}\left(\delta_{\sigma,\sigma_{i}}-\delta_{\sigma,\sigma_{j}}\right).\end{eqnarray*}
 So $P_{ij}$ acts as a translation operator, and when acting on a general bosonic wavefunction $\chi\left(\phi_{n}^{\sigma}\right)$, we have\begin{equation}
 P_{ij}\chi\left(\phi_{n}^{\sigma}\right)=e^{\left\{
P\left(x_{i},\sigma_{i};x_{j},\sigma_{j}\right)\right\}}\chi\left(\phi_{n}^{\sigma}\right)\,.\label{Applying
Pij to sym wavefctn}\end{equation}
 where\begin{eqnarray}
P\left(x_{i},\sigma_{i};x_{j},\sigma_{j}\right) & \equiv & \sum_{\sigma}\left(\delta_{\sigma\sigma_{i}}-\delta_{\sigma\sigma_{j}}\right)\left(\Pi^{\sigma}\left(x_{i}\right)-\Pi^{\sigma}\left(x_{j}\right)\right)\nonumber\\
 & = & \int_{x_{j}}^{x_{i}}dx\left(\partial_{x}\Pi^{\sigma_{i}}\left(x\right)-\partial_{x}\Pi^{\sigma_{j}}\left(x\right)\right).\end{eqnarray}

Finally, $H_{s}$ can be rewritten as%
\footnote{In order to write this and other sums that diverge when $i=j$, we
use the following result. Let $f\left(x_{i},x_{j}\right)$ be a function
that describes some 2-body interaction, either singular or not when
$i\rightarrow j$. Then\begin{eqnarray*}
 \sum_{i\ne j}f\left(x_{i},\sigma_{i};x_{j},\sigma_{j}\right) & = & \sum_{i\ne j}\sum_{\sigma,\sigma'}\int dx\delta_{\sigma\sigma_{i}}\delta_{\sigma'\sigma_{j}}\delta\left(x-x_{i}\right)\int dy\delta\left(y-x_{j}\right)f\left(x,\sigma;y,\sigma'\right)\\
  & = & \sum_{\sigma,\sigma'}\int dx\int dyf\left(x,\sigma;y,\sigma'\right)\left\{ \sum_{i,j}-\sum_{i=j}\right\} \delta_{\sigma\sigma_{i}}\delta_{\sigma'\sigma_{j}}\delta\left(x-x_{i}\right)\delta\left(y-x_{j}\right)\\
  & = & \sum_{\sigma,\sigma'}\int dx\int dy\phi_{\sigma}\left(x\right)f\left(x,\sigma;y,\sigma'\right)\left\{ \phi_{\sigma'}\left(y\right)-\delta_{\sigma\sigma'}\delta\left(y-x\right)\right\} .\end{eqnarray*}
 This result allows us to regularize the integrals, once we use the
density formulation. One other way of writing this result is by the
use of principal value integrals:\[ \sum_{i\ne
j}f\left(x_{i},\sigma_{i};x_{j},\sigma_{j}\right)=\sum_{\sigma,\sigma'}\int
dx\dashint
dy\phi_{\sigma}\left(x\right)f\left(x,\sigma;y,\sigma'\right)\phi_{\sigma'}\left(y\right).\]
}\begin{eqnarray}
 H_{s} & = & -\beta\sum_{i\ne j}\frac{z_{i}z_{j}}{\left(z_{i}-z_{j}\right)^{2}}\left(P_{ij}-1\right)\nonumber \\
  & = & \frac{\beta}{4}\sum_{\sigma,\sigma'}\int dx\int dy\phi_{\sigma}\left(x\right)\sin^{-2}\left(\frac{x-y}{2}\right)\left\{ \phi_{\sigma'}\left(y\right)-\delta_{\sigma\sigma'}\delta\left(y-x\right)\right\} \left\{ e^{P\left(x,\sigma;y,\sigma'\right)}-1\right\} \nonumber \\
  & = & \frac{\beta}{4}\sum_{\sigma,\sigma'}\int dx\int dy\phi_{\sigma}\left(x\right)\sin^{-2}\left(\frac{x-y}{2}\right)\phi_{\sigma'}\left(y\right)\left\{ \sum_{n=1}^{\infty}\frac{1}{n!}\left[\int_{y}^{x}dz\partial_{z}\left(\Pi^{\sigma}\left(z\right)-\Pi^{\sigma'}\left(z\right)\right)\right]^{n}\right\} .\nonumber
 \\  \label{Define Hs}
\end{eqnarray}
 Note that this interaction term vanishes for $\sigma=\sigma'$, so
there is no issue at $y=x$. In fact the sum over both $\sigma,\sigma'$
could be substituted by a sum over $\sigma\ne\sigma'$.

To complete the comparison with \cite{Awata:1995by}, we still need
to confirm this last term of the hamiltonian, $H_{s}$.

As it was seen in (\ref{Applying Pij to sym wavefctn}), the operator
$P_{ij}$ acts on bosonic wave functions as:\begin{equation*}
e^{P\left(x,\sigma;y,\sigma'\right)} = \exp\left\{
\sum_{n}\frac{1}{n}\left(z_{x}^{n}-z_{y}^{n}\right)\left(\hat{\Pi}_{n}^{\sigma}-\hat{\Pi}_{n}^{\sigma'}\right)\right\}
=:\left(P_{x}^{\sigma\sigma'}\right)\left(P_{y}^{\sigma\sigma'}\right)^{-1}:\,,\end{equation*}
 where \[
P_{x}^{\sigma\sigma'}=\exp\left\{
\sum_{n}\frac{1}{n}z_{x}^{n}\left(\hat{\Pi}_{n}^{\sigma}-\hat{\Pi}_{n}^{\sigma'}\right)\right\}
=\exp\left\{
\Pi^{\sigma}\left(z_{x}\right)-\Pi^{\sigma'}\left(z_{x}\right)\right\}
,\]
 and $z_{x}\equiv e^{ix}$. If we go to complex coordinates,%
\footnote{In order to go to complex coordinates, remember that
$\phi_{\sigma}\left(x\right)$ has conformal weight one, much like
$\psi^{\dagger}$ in the previous
section, and so transforms like $\phi_{\sigma}\left(x\right)\rightarrow z\phi_{\sigma}\left(z\right)$.%
} we get (as before, we have to be careful in the integration, as
the integration over $w$ will be around the coordinate
$z$)\begin{eqnarray*}
 H_{s} & = & \beta\sum_{\sigma\ne\sigma'}\oint dz\oint dw\frac{zw}{\left(z-w\right)^{2}}\phi_{\sigma}\left(z\right)\phi_{\sigma'}\left(w\right)\left\{ e^{P\left(z,\sigma;w,\sigma'\right)}-1\right\} \\
  & = & \beta\sum_{\sigma\ne\sigma'}\oint dz\oint dw\frac{1}{\left(z-w\right)^{2}}:z\phi_{\sigma}\left(z\right)e^{\Pi^{\sigma}\left(z\right)-\Pi^{\sigma'}\left(z\right)}w\phi_{\sigma'}\left(w\right)e^{-\Pi^{\sigma}\left(z\right)+\Pi^{\sigma'}\left(z\right)}:\,.\end{eqnarray*}
 Some simplifications in the above expression were due to restricting the expansion
of $\phi$ to positive modes:
$\phi_{\sigma}\left(z\right)=\sum_{n\geq0}\phi_{n}^{\sigma}z^{-n-1}$.
There is one other way of writing this result:
\begin{eqnarray*}
 H_{s} & = & \beta\sum_{\sigma\ne\sigma'}\oint\frac{d\eta}{\eta}\oint\frac{d\xi}{\xi}\sum_{k,\ell\geq0}\eta^{k}\xi^{\ell}\int dx\phi_{\sigma}\left(x\right)z_{x}^{k}\int dy\phi_{\sigma'}\left(y\right)z_{y}^{\ell}\frac{z_{x}z_{y}e^{\sum_{n>0}\frac{1}{n}\left(\eta^{-n}-\xi^{-n}\right)\left(\hat{\Pi}_{n}^{\sigma}-\hat{\Pi}_{n}^{\sigma'}\right)}}{\left(z_{x}-z_{y}\right)^{2}}\\
  & = & \frac{\beta}{2}\sum_{\sigma\ne\sigma'}\oint\frac{d\eta}{\eta}\oint\frac{d\xi}{\xi}\sum_{k,\ell\geq0}\eta^{k}\xi^{\ell}\phi_{k}^{\sigma}\phi_{\ell}^{\sigma'}e^{\sum_{n>0}\frac{1}{n}\left(\eta^{-n}-\xi^{-n}\right)\left(\hat{\Pi}_{n}^{\sigma}-\hat{\Pi}_{n}^{\sigma'}\right)}\sum_{m>0}m\left[\left(\frac{\eta}{\xi}\right)^{m}+\left(\frac{\xi}{\eta}\right)^{m}\right].\end{eqnarray*}

These results represent the collective field bosonization of the C-S
many body problem. They are in accordance with the expressions found
in \cite{Awata:1995by}. One can further ask regarding  a connection
with the bosonization procedure of section 3. In one case we dealt
with a fermionic effective hamiltonian and followed standard rules
of bosonization. In the other one first went to a bosonic picture
and then applied the technique of collective fields. Since the
relation between the fermionic and the bosonic pictures can be
described in terms of a single similarity transformation it is
expected that the two versions of the continuum hamiltonian are
equivalent. We can expect that the identification can be
demonstrated by a field theoretic similarity transformation. Its
explicit form remains to be understood.

To conclude this section, we now present the total collective
hamiltonian corresponding to the spin Calogero-Sutherland model, in
coordinate space:\begin{eqnarray}
 H & = & E+\beta\sum_{\sigma}\int dx\,\phi_{\sigma}\left(x\right)\partial^{2}\Pi^{\sigma}\left(x\right)-\sum_{\sigma}\int dx\phi_{\sigma}\left(x\right)\partial\Pi^{\sigma}\left(x\right)\partial\Pi^{\sigma}\left(x\right)-\nonumber \\
  &  & \qquad\qquad-\left(\beta+1\right)\sum_{\sigma,\sigma'}\int dx\phi_{\sigma}\left(x\right)\int dy\phi_{\sigma'}\left(y\right)\cot\left(\frac{x-y}{2}\right)\partial\Pi^{\sigma}\left(x\right)+\nonumber \\
  &  & +\frac{\beta}{4}\sum_{\sigma,\sigma'}\int dx\int dy\frac{\phi_{\sigma}\left(x\right)\phi_{\sigma'}\left(y\right)}{\sin^{2}\left(\frac{x-y}{2}\right)}\left\{ \sum_{n=1}^{\infty}\frac{1}{n!}\left[\int_{y}^{x}dz\partial_{z}\left(\Pi^{\sigma}\left(z\right)-\Pi^{\sigma'}\left(z\right)\right)\right]^{n}\right\} .\label{1 - Full Hamiltonian}\end{eqnarray}

The hamiltonian thus presented is comparable to what was obtained by
the bosonization procedure of section \ref{Current-Algebra Rep},
given by (\ref{Bosonized Hamiltonian}), and will be used in the
following sections.

\setcounter{footnote}{0}

\section{Spectrum Equation}\label{M-O Application}

The next two sections will be devoted to the study of some
applications of the bosonized spin CS hamiltonian. In this section,
we will take a second look at the collective form of the
hamiltonian, and put it in terms of spin and charge bosons, as was
done in section \ref{Current-Algebra Rep} of the paper. In this form
of the hamiltonian, we can use semi-classical method of determining
the energies, in order to be able to perform small fluctuations.This
will result in an eigenvalue problem of  the Marchesini-Onofri type.
Then, in section \ref{Nontrivial Bckgrd}, we will describe the
formalism in an example given by a nontrivial background based on
exact eigenstates of \cite{ha:1992-}.

\subsection{The Marchesini-Onofri kernel}

We now go back to the effective hamiltonian $H_{\beta+1}$, in its
collective form (\ref{1 - Full Hamiltonian}), but will regularize
the integrals by the use of principal value integrals. It is easy to
check that, this be the case, (\ref{1 - Full Hamiltonian}) can be
rewritten as:\begin{eqnarray*}
 H & = & E-\sum_{\sigma}\int dx\,\phi_{\sigma}\left(x\right)\partial^{2}\Pi^{\sigma}\left(x\right)-\sum_{\sigma}\int dx\phi_{\sigma}\left(x\right)\partial\Pi^{\sigma}\left(x\right)\partial\Pi^{\sigma}\left(x\right)-\\
  &  & \qquad\qquad-\left(\beta+1\right)\sum_{\sigma,\sigma'}\int dx\phi_{\sigma}\left(x\right)\dashint dy\phi_{\sigma'}\left(y\right)\cot\left(\frac{x-y}{2}\right)\partial\Pi^{\sigma}\left(x\right)+\\
  &  & \frac{\beta}{4}\sum_{\sigma\ne\sigma'}\int dx\dashint dy\phi_{\sigma}\left(x\right)\sin^{-2}\left(\frac{x-y}{2}\right)\phi_{\sigma'}\left(y\right)\left\{ e^{P\left(x^{\sigma},y^{\sigma'}\right)}-1\right\} ,\end{eqnarray*}
 with $E=\left(\beta+1\right)^{2}\frac{N\left(N^{2}-1\right)}{12}$
and \begin{equation}
P\left(x,\sigma;y,\sigma'\right)=\int_{y}^{x}dz\partial_{z}\left(\Pi^{\sigma}\left(z\right)-\Pi^{\sigma'}\left(z\right)\right)=\left(\Pi^{\sigma}-\Pi^{\sigma'}\right)\left(x\right)-\left(\Pi^{\sigma}-\Pi^{\sigma'}\right)\left(y\right).\end{equation}

 The next step is to introduce the spin and charge bosons, just like
we did in section \ref{Current-Algebra Rep}. Remembering that we are
dealing with an $su\left(2\right)$ current algebra, let\[
\phi=\sum_{\sigma}\phi_{\sigma}\qquad;\qquad\psi=\sum_{\sigma}\sigma\phi_{\sigma}.\]
 Then, as a consequence of
 $\Pi_{\sigma}\equiv\frac{\partial}{\partial\phi_{\sigma}}$,
we have\[
\Pi_{\phi}\equiv\frac{\delta}{\delta\phi}=\sum_{\sigma}\Pi_{\sigma}\quad;\quad\Pi_{\psi}\equiv\frac{\delta}{\delta\psi}=\sum_{\sigma}\sigma\Pi_{\sigma}.\]

 With this definitions we can rewrite the hamiltonian in terms of
the new fields (by using that
$\phi_{\sigma}=\frac{1}{2}\left(\phi+\sigma\psi\right)$ and
$\Pi_{\sigma}=\frac{1}{2}\left(\Pi_{\phi}+\sigma\Pi_{\psi}\right)$).
Recalling equations (\ref{Define H1}), (\ref{Define H21 H22}) and
(\ref{Define Hs}) one can easily obtain the following
results:\begin{eqnarray*}  H_{1} & = & -\frac{1}{2}\int
dx\left(\partial^{2}\phi\Pi_{\phi}+\partial^{2}\psi\Pi_{\psi}\right)-\frac{1}{2}\int dx\dashint dy\delta\left(x-y\right)\psi\left(x\right)\partial\Pi_{\phi}\left(x\right)\partial\Pi_{\psi}\left(y\right) \\
  &  & -\frac{1}{4}\int dx\dashint dy\delta\left(x-y\right) \phi\left(x\right)\left[\partial\Pi_{\phi}\left(x\right)\partial\Pi_{\phi}\left(y\right)+\partial\Pi_{\psi}\left(x\right)\partial\Pi_{\psi}\left(y\right)\right] ,\\
 H_{2} & = & \frac{\beta+1}{2}\int
dx\left(\partial^{2}\phi\Pi_{\phi}+\partial^{2}\psi\Pi_{\psi}\right)\\
  &  & -\frac{\beta+1}{2}\int dx\dashint dy\phi\left(y\right)\cot\left(\frac{x-y}{2}\right)\left[\phi\left(x\right)\partial\Pi_{\phi}\left(x\right)+\psi\left(x\right)\partial\Pi_{\psi}\left(x\right)\right].\end{eqnarray*}
 The only term left to determine is $H_{s}$. Because we only have
two spins ($\sigma=\pm1$), the sum over $\sigma\ne\sigma'$ in
$H_{s}$ can be substituted by a sum $\sigma>\sigma'$ (which is not
really a sum, as the only possibility is $\sigma=1,\sigma'=-1$).
Then using\[
P\left(x,\uparrow;y,\downarrow\right)=\left(\Pi^{\uparrow}-\Pi^{\downarrow}\right)\left(x\right)-\left(\Pi^{\uparrow}-\Pi^{\downarrow}\right)\left(y\right)=\Pi_{\psi}\left(x\right)-\Pi_{\psi}\left(y\right),\]
 we can write $H_{s}$ in the following way:\[
 H_{s}=\frac{\beta}{2}\sum_{\sigma>\sigma'}\int dx\dashint
dy\phi_{\uparrow}\left(x\right)\phi_{\downarrow}\left(y\right)\sin^{-2}\left(\frac{x-y}{2}\right)\left\{
e^{\Pi_{\psi}\left(x\right)-\Pi_{\psi}\left(y\right)}-1\right\}
=H_{s}^{1}+H_{s}^{2},\]
 where after some algebra one gets:\begin{eqnarray}
 H_{s}^{1} & = & -\frac{\beta}{8}\int dx\dashint dy\frac{\left[\phi\left(x\right)\phi\left(y\right)-\psi\left(x\right)\psi\left(y\right)\right]}{\sin^{2}\left(\frac{x-y}{2}\right)}\left\{ 1-e^{\Pi_{\psi}\left(x\right)-\Pi_{\psi}\left(y\right)}\right\} ,\nonumber \\
 H_{s}^{2} & = & -\frac{\beta}{4}\int dx\dashint
dy\frac{\phi\left(x\right)\psi\left(y\right)}{\sin^{2}\left(\frac{x-y}{2}\right)}\sinh\left[\Pi_{\psi}\left(x\right)-\Pi_{\psi}\left(y\right)\right].\label{Hs
in terms of fields phi and psi}\end{eqnarray}

 The first conclusion that one can draw is that the interaction term
$H_{s}$ does \textbf{not} depend on the field $\Pi_{\phi}$. This
means that the full hamiltonian will be at most quadratic in this
field, and we can proceed to hermitianize the hamiltonian in terms
of the fields $\phi,\Pi_{\phi}$, following the standard procedure
shown in \cite{Jevicki:1979mb,Jevicki:1980zg}. To do so, let us
first re-write the full hamiltonian in a simpler
way:\begin{eqnarray}
 H & = & \int dx\,\omega^{\phi}\left(x\right)\Pi_{\phi}+\int dx\int dy\Omega_{x,y}^{\phi}\Pi_{\phi}\left(x\right)\Pi_{\phi}\left(y\right)+2\int dx\int dy\widetilde{\Omega}_{x,y}\Pi_{\phi}\left(x\right)\Pi_{\psi}\left(y\right)+\nonumber \\
  &  & +\int dx\,\omega^{\psi}\left(x\right)\Pi_{\psi}+\int dx\int dy\Omega_{x,y}^{\psi}\Pi_{\psi}\left(x\right)\Pi_{\psi}\left(y\right)+H_{s}+E,\label{spin/charge collective hamiltonian}\end{eqnarray}
 where the definitions for these terms are given below:\begin{eqnarray}
\omega^{\phi}\left(x\right) & = & -\frac{1}{2}\partial^{2}\phi\left(x\right)+\frac{\beta+1}{2}\partial_{x}\left[\phi\left(x\right)\dashint dy\phi\left(y\right)\cot\left(\frac{x-y}{2}\right)\right];\\
\omega^{\psi}\left(x\right) & = & -\frac{1}{2}\partial^{2}\psi\left(x\right)+\frac{\beta+1}{2}\partial_{x}\left[\psi\left(x\right)\dashint dy\phi\left(y\right)\cot\left(\frac{x-y}{2}\right)\right];\\
\Omega_{x,y}^{\phi} & = & -\frac{1}{4}\partial_{x}\partial_{y}\left[\delta\left(x-y\right)\phi\left(x\right)\right];\\
\Omega_{x,y}^{\psi} & = & -\frac{1}{4}\partial_{x}\partial_{y}\left[\delta\left(x-y\right)\phi\left(x\right)\right];\\
\widetilde{\Omega}_{x,y} & = & -\frac{1}{4}\partial_{x}\partial_{y}\left[\delta\left(x-y\right)\psi\left(x\right)\right].\end{eqnarray}

 As we said, we are interested in hemitianizing the hamiltonian with
respect to the fields $\phi,\Pi_{\phi}$. We know from
\cite{Jevicki:1979mb,Jevicki:1980zg} that it can be done by a change
of variables, equivalent to the similarity transformation\[
\phi\rightarrow\phi\quad;\qquad\Pi_{\phi}\rightarrow\Pi_{\phi}-\frac{1}{2}\frac{\partial\ln
J}{\partial\phi},\]
 where the Jacobian $J$ obeys\begin{equation}
\frac{\partial\ln J}{\partial\phi}=\int
dy\left(\Omega^{\phi}\right)^{-1}_{x,y}\omega^{\phi}\left(y\right).\end{equation}
 With this transformation, the first two terms in (\ref{spin/charge collective hamiltonian}),
i.e. the terms that have no $\Pi_{\psi}$ dependence, become%
\footnote{It is easy to show that $
\partial_{x}\partial_{y}\left(\Omega_{x,y}^{\phi}\right)^{-1}=-\frac{4}{\phi\left(x\right)}\delta\left(x-y\right).$
}\begin{eqnarray}
 H_{\phi} & \rightarrow & -\frac{1}{4}\int dx\partial\Pi_{\phi}\left(x\right)\phi\left(x\right)\partial\Pi_{\phi}\left(x\right)+\frac{\left(\beta+1\right)}{2}\int dx\phi\left(x\right)\partial_{x}\left[\dashint dy\phi\left(y\right)\cot\left(\frac{x-y}{2}\right)\right]+\nonumber \\
  &  & \qquad-\frac{N}{12}\left(\beta+1\right)^{2}+\frac{\pi^{2}}{3}\left(\beta+1\right)^{2}\int dx\phi^{3}\left(x\right)+\frac{\beta^{2}}{4}\int
 dx\frac{\left(\partial\phi\right)^{2}}{\phi},\label{Hermitian part of hamiltonian}\end{eqnarray}
 where we have used the following
identity (which can be proven by Fourier transform):\begin{equation}
\frac{1}{4}\int dx\phi\left(x\right)\left[\dashint dy\phi\left(y\right)\cot\left(\frac{x-y}{2}\right)\right]^{2}=\frac{\pi^{2}}{3}\int dx\phi^{3}\left(x\right)-\frac{N^{3}}{12}.\label{Fourier identity}\end{equation}

But these terms are not the only ones that get changed with the
transformation. The third term of (\ref{spin/charge collective
hamiltonian}) also gets shifted, as it has a field $\Pi_{\phi}$:
\begin{eqnarray*}
 H_{\phi\psi} & \rightarrow & 2\int dx\int dy\widetilde{\Omega}_{x,y}\left(\Pi_{\phi}\left(x\right)-\frac{1}{2}\dashint dz\left(\Omega_{x,z}^{\phi}\right)^{-1}\omega^{\phi}\left(z\right)\right)\Pi_{\psi}\left(y\right)\\
  & = & 2\int dx\int dy\widetilde{\Omega}_{x,y}\Pi_{\phi}\left(x\right)\Pi_{\psi}\left(y\right)+\\
  &  & +\frac{1}{2}\int dx\partial_{x}\left\{ \psi\left(x\right)\frac{\left(\partial\phi\right)}{\phi}-\left(\beta+1\right)\psi\left(x\right)\int dy\phi\left(y\right)\cot\left(\frac{x-y}{2}\right)\right\} \Pi_{\psi}\left(y\right).\end{eqnarray*}
 All the other terms in the hamiltonian do not change under the transformation.
The total hamiltonian after the transformation
becomes:\begin{equation}  \tilde{H}=H_{\phi}+2\int\! dx\!\int\!
dy\widetilde{\Omega}_{x,y}\Pi_{\phi}\left(x\right)\Pi_{\psi}\left(y\right)+\int
dx\,\tilde{\omega}^{\psi}\left(x\right)\Pi_{\psi}+\int\! dx\!\int\!
dy\Omega_{x,y}^{\psi}\Pi_{\psi}\left(x\right)\Pi_{\psi}\left(y\right)+H_{s},\end{equation}
 where we have defined all the quantities above, except for the new $\tilde{\omega}^{\psi}$,
given by\begin{equation}
\tilde{\omega}^{\psi}\left(x\right)=\frac{1}{2}\partial_{x}\left(\psi\frac{\partial_{x}\phi}{\phi}-\partial_{x}\psi\right).\end{equation}

The objective of such a transformation is to be able to study
fluctuations about a single matrix background, which is a stationary
point of the explicitly hermitian effective potential in the fields
$\phi$. As we will be interested in small fluctuations around this
stationary point, we can still expand the term $H_{s}$ of the
hamiltonian up to third order in the fields (ignoring higher order
effects). Keeping only linear, quadratic and cubic terms, we have
$H_{s}^{1} \approx 0$, and so
\begin{eqnarray}
H_{s} & \approx & -\frac{\beta}{4}\int dx\dashint
dy\frac{\left[\phi\left(x\right)\psi\left(y\right)-\psi\left(x\right)\phi\left(y\right)\right]}{\sin^{2}\left(\frac{x-y}{2}\right)}\Pi_{\psi}\left(x\right).\label{Expansion
of interaction hamiltonian}\end{eqnarray}
 We still want to make one other assumption. We will assume that the
spin field $\psi$ does not acquire an expectation value in the
classical equations of motion, and so we will drop the term with
double $\Pi_{\psi}\Pi_{\psi}$ in the hamiltonian, $H_{\psi\psi}$.
The reason for this is that this term is directly related to the
term of current-current interaction (the $J^{2}$ term in section
\ref{Current-Algebra Rep}). In fact $\psi$ was seen to be the
bosonic field used to bosonize the current terms (in section
\ref{Current-Algebra Rep}, we called it $\beta$) and as the current
interaction will be zero at the classical stationary point, the term
$H_{\psi\psi}$ of the hamiltonian will not be included in the final
version of the hamiltonian.

Finally we write the full hamiltonian as the sum of \[
H=H_{\phi}+\Delta H,\]
 with \begin{eqnarray}
 \Delta H & = & -\frac{1}{2}\int dx\partial\Pi_{\phi}\left(x\right)\psi\left(x\right)\partial\Pi_{\psi}\left(x\right)+\\
  &  & +\frac{1}{2}\int
dx\partial_{x}\left(\psi\frac{\partial_{x}\phi}{\phi}-\partial_{x}\psi\right)\Pi_{\psi}-\frac{\beta}{4}\int
dx\dashint
dy\frac{\left[\phi\left(x\right)\psi\left(y\right)-\psi\left(x\right)\phi\left(y\right)\right]}{\sin^{2}\left(\frac{x-y}{2}\right)}\Pi_{\psi}\left(x\right).\nonumber\end{eqnarray}

 As mentioned before, our intention is to consider small
 fluctuations around a classical background field configuration.
 The term $H_{\phi}$ of the hamiltonian allows us, in the large $N$
limit, to determine the classical background
$\phi\left(x\right)=\phi_{0}\left(x\right)$, with $\psi_{0}=0$
(solving the variation associated with $H_{\phi}$, and ignoring
several terms which are higher order in $1/N$
\cite{Andric:1982jk,Jevicki:1991yi,Abanov:2005nt}).%
 Also, to obtain $\phi_{0}$ one imposes that $\Pi_{\psi}$ doesn't
acquire a classical expectation value. The ground state thus
obtained has $M_{\downarrow}=M_{\uparrow}=\frac{N}{2}$ (as was said,
$\psi$ will also have no classical expectation value), and
$J_{\downarrow}=J_{\uparrow}=J$.

To obtain energies above the ground state, we need to allow $\psi$
and $\Pi_{\psi}$ to have non-zero expectation value, and solve for
the full set of equations of motion. But  allowing $\Pi_{\psi}$ to
have a non-zero expectation value invalidates the expansion
(\ref{Expansion of interaction hamiltonian}). We would have to use
the full expression (\ref{Hs in terms of fields phi and psi}), and
only perform an expansion once we have the classical values for the
fields. That expansion would then truncate by use of the theorems
found in \cite{ha:1992-}. These theorems allow us to determine the
energy contribution of the spin-interaction term in the hamiltonian,
in particular proving that the higher spin states do not contribute.

To study the theory perturbatively, as long as we consider that
$\Pi_{\psi}$ does not acquire an expectation value, we can use the
results obtained in this section.

\subsection{Non-critical Strings}

In applications to both critical and noncritical strings, linearized
eigenvalue problems of this kind appear. In recent studies of the
AdS/CFT correspondence \cite{Donos:2005vm} a similar equation is
seen to determine the spectrum of BPS states. The eigenvalue problem
itself plays a role in the nontrivial map between the matrix model
and AdS space-time.

While the regular harmonic oscillator participates in the $1/2$ BPS
map, for the 2d noncritical string one requires an inverted
oscillator potential. So, for the noncritical strings, the classical
stationary solution (with  $\Pi_{\phi}\equiv0$) is given by
$\phi_{0}\left(x\right)\propto\sqrt{x^{2}-\mu}$
 \cite{Jevicki:1993qn,Klebanov:1991qa,Gross:1990md,Avan:1991ik}.

In the background of the classical value for the field $\phi$, we
will now want to study the interaction part of the hamiltonian:
\begin{eqnarray}
 \left.\Delta H\right|_{\phi_{0}} & = & -\frac{\beta}{4}\int dx\dashint dy\frac{\left[\phi_{0}\left(x\right)\psi\left(y\right)-\psi\left(x\right)\phi_{0}\left(y\right)\right]}{\sin^{2}\left(\frac{x-y}{2}\right)}\Pi_{\psi}\left(x\right)\nonumber \\
  & = & \int dx\dashint dy\psi\left(x\right)K\left(x,y\right)\Pi_{\psi}\left(x\right),\label{Hamiltonian at stationary point}\end{eqnarray}
 where we have \begin{equation}
 \int dx\psi\left(x\right)\int
dyK\left(x,y\right)\Pi_{\psi}\left(x\right)=-\frac{\beta}{4}\int
dx\psi\left(x\right)\int
dy\phi_{0}\left(y\right)\frac{\Pi_{\psi}\left(y\right)-\Pi_{\psi}\left(x\right)}{\sin^{2}\left(\frac{x-y}{2}\right)}.\label{Marchesini-Onofri
kernel}\end{equation}
 which acts on wave functionals%
\footnote{Recall that the fields $\phi_{\sigma}$ have to obey the
following normalization\[ \int
dx\phi_{\sigma}\left(x\right)=M_{\sigma}\quad\rm{and}\qquad\sum_{\sigma}M_{\sigma}=N.\]
 Furthermore, if the fields $\Pi_{\sigma}$ have a classical expectation
value of $J_{\sigma}$, in coordinate space we'll have\[ \int
dx\phi_{\sigma}\left(x\right)\partial\Pi_{\sigma}\left(x\right)=M_{\sigma}J_{\sigma}.\]
}
 $\Phi=\int dzf\left(z\right)\psi\left(z\right).$ This
$-K\left(x,y\right)$ is the Marchesini-Onofri kernel
\cite{Marchesini:1979yq,Gross:1990md}.  We conclude that the
Marchesini-Onofri kernel comes directly from the spin-interaction
term of the hamiltonian.

We are interested in solving the eigenvalue problem $\left(\Delta
H\right)\Phi=\epsilon \Phi$, which can be written in a simpler way.
Consider a complete set of eigenfunctions of $K\left(x,y\right)$,
$\left\{ f_{n}\left(x\right)\right\} $, with normalization condition%
\[ \int
dx\phi_{0}\left(x\right)f_{n}\left(x\right)f_{m}\left(x\right)=\delta_{mn}.\]
 Expand the fields $\psi,\Pi_{\psi}$ in this
basis:\begin{equation*} \Pi_{\psi}\left(x\right)  =
\phi_{0}\left(x\right)\sum_{n}f_{n}\left(x\right)\Pi_{n}\quad;\qquad
\psi\left(x\right)  =
\sum_{n}f_{n}\left(x\right)\psi_{n}.\end{equation*}%
 With these expansions, we can write $\Delta H$ as \[ \Delta
H=\sum_{m,n}\psi_{n}\Pi_{m}\int dxf_{n}\left(x\right)\dashint
dyK\left(x,y\right)\phi_{0}\left(y\right)f_{m}\left(y\right)=\sum_{n}\omega_{n}\psi_{n}\Pi_{n},\]
where the frequencies $\omega_{n}$ are defined by the following eigenvalue equation%
\begin{equation} \dashint
dy\phi_{0}\left(y\right)K\left(x,y\right)f_{m}\left(y\right)=\omega_{m}\phi_{0}\left(x\right)f_{m}\left(x\right).\end{equation}
One can use the semiclassical results found in \cite{Gross:1990md}
to evaluate these frequencies, as the eigenvalue problems found in
each case are equivalent. The result is then:
\[
\omega_{n}=\epsilon_{n}=\left(n+1\right)\omega\left(\mu\right)+\eta\left(\mu\right),\]
with
$\omega\left(\mu\right)=\frac{1}{\eta\left(\mu\right)}=\frac{2\beta}{\left|\ln\mu\right|}$.
 The term  $\omega\left(\mu\right)$ measures the splitting in the singlet
energies. These splittings tend to zero in the $\mu\rightarrow0$
limit (double scaling limit). The term $\eta\left(\mu\right)$ gives
the difference in energy between the singlet state (corresponding to
trivial representation, if we set the interaction term to zero) and
the adjoint state due to the interaction term. Up to leading order,
$\eta\left(\mu\right)$ can be identified with the ground state
energy $\omega_{0}$, divergent in the double scaling limit
\cite{Gross:1990md}.

In \cite{Maldacena:2005hi} the same eigenvalue problem was studied,
and it was shown to be equivalent to\begin{equation}
\hat{\epsilon}h\left(\tau\right)=-\frac{1}{\pi}\int_{-\infty}^{\infty}\frac{h\left(\tau'\right)}{4\sinh^{2}\left(\frac{\tau-\tau'}{2}\right)}+\hat{v}\left(\tau\right)h\left(\tau\right),\end{equation}
with
$\hat{v}\left(\tau\right)=-\frac{1}{\pi}\frac{\tau}{\tanh\tau}$, and
$\hat{\epsilon}=\epsilon+\frac{\ln\mu}{2\pi}$, where $\epsilon$ is
the energy in the original eigenvalue problem,
$h\left(\tau\right)=\phi_{0}\left(\tau\right)f\left(\tau\right)$,
and $\tau$ is the time-of-flight coordinate:
$x=\sqrt{2\mu}\cosh\tau$. In the energy $\hat{\epsilon}$ we
subtracted the divergent piece,
$\epsilon_{0}\approx\eta\left(\mu\right)$, renormalizing the energy.

The comparison to string theory allows us to interpret this
divergent piece of the energy. To perform this comparison, valid for
large $\tau$, we have to identify $\tau\sim-\varphi$, where
$\varphi$ is the Liouville direction \cite{Ginsparg:1993is}. The
divergence that leads to the renormalization of the energy is then
basically related with the fact that the string is stretching from
the Liouville wall to $-\infty$.

Also in \cite{Maldacena:2005hi}, a scattering phase (in the
background of the inverted oscillator potential) was conjectured to
be:\begin{equation}
\delta\left(\hat{\epsilon}\right)=-\int_{-\infty}^{\infty}\left(\frac{\pi\epsilon'}{\tanh\pi\epsilon'}+\pi\epsilon'\right)d\epsilon',\end{equation}
as it is in agreement with the scattering phase of the Liouville
model, in the string theory approach. This result was seen to come
from solving the Marchesini-Onofri eigenvalue problem, first for the
asymptotic limits $\hat{\epsilon}\ll0$ and $\hat{\epsilon}\gg0$
\cite{Maldacena:2005hi}, and later for any energy
\cite{Fidkowski:2005ck}.

\section{Nontrivial Eigenstates and Backgrounds}\label{Nontrivial Bckgrd}

 Of major interest in studies and applications of matrix models are possibilities of
introducing non-trivial backgrounds in the theory. These are in
correspondence with nontrivial states of the many body problem. In
the case of free fermions these were for example states representing
particles and holes and associated solitonic configurations
\cite{Jevicki:1991yi,Andric:1996nn}. In the present case of
Spin-Calogero problem an interesting class of nontrivial exact
eigenstates was given by Ha and Haldane in \cite{ha:1992-} . They
constructed a many body wavefunction of the form is given by
\begin{equation} \Psi=\Delta^{\beta}\tilde{\Delta}\chi,\label{1 -
wavefunction ansatz}\end{equation}
 where \begin{eqnarray}
\Delta & = & \prod_{i<j}\left(z_{i}-z_{j}\right)\prod_{i}z_{i}^{-\frac{N-1}{2}}\nonumber \\
\widetilde{\Delta} & = & \prod_{i<j}\left(z_{i}-z_{j}\right)^{\delta_{\sigma_{i}\sigma_{j}}}e^{i\frac{\pi}{2}\rm{sgn}\left(\sigma_{i}-\sigma_{j}\right)},\\
\chi\left(\phi_{n}\right) & = &
\Pi_{i}z_{i}^{J_{\sigma_{i}}}=e^{i\sum_{j}J_{\sigma_{j}}x_{j}}=e^{i\sum_{\sigma}J_{\sigma}\int
dx\, x\phi^{\sigma}\left(x\right)}.\label{1 - Wavefunction ansatz
2}\end{eqnarray}
and evaluated the corresponding energy.

It is our interest to consider this state in collective field
theory. The strategy is to perform a similarity transformation, but
now in terms of the new wavefunction
\begin{eqnarray}
\Psi_{0} & = & \Delta^{\beta}\widetilde{\Delta}\,.\label{2nd Form of the wave function}\end{eqnarray}
 The resulting effective hamiltonian describes the theory above the nontrivial quantum state. It can be written as a sum of several
terms:\begin{eqnarray}
 \overline{H} & \equiv & \Psi_{0}^{-1}H\Psi_{0}\label{Haldane effective hamiltonian}\\
  & = & \underbrace{\sum_{i}D_{i}^{2}}_{\overline{H}_{1}}+\underbrace{\beta\sum_{i<j}\left(\theta_{ij}-\theta_{ji}\right)\left(D_{i}-D_{j}\right)}_{\overline{H}_{2}}+\underbrace{2\sum_{i\ne j}\theta_{ij}\delta_{\sigma_{i}\sigma_{j}}D_{i}}_{\overline{H}_{3}}+\overline{H}_{int}+\overline{H}_{local}+E\,,\nonumber \end{eqnarray}

 There are two contributions to the ground state energy, one local,
$\overline{H}_{local}$), and the other non local, $E$. They can be
 determined to be:\begin{eqnarray}
E & = & E_{0}+\frac{1}{2}\left(\beta+1\right)\sum_{\sigma}M_{\sigma}\left(M_{\sigma}-1\right)+\frac{1}{3}\sum_{\sigma}M_{\sigma}\left(M_{\sigma}-1\right)\left(M_{\sigma}-2\right),\\
 \overline{H}_{local} & = & \beta\sum_{i\ne j\ne
k}\frac{\left(z_{i}+z_{j}\right)z_{i}\delta_{\sigma_{i}\sigma_{k}}}{\left(z_{i}-z_{j}\right)\left(z_{i}-z_{k}\right)}+\beta\sum_{i\ne
j}\frac{z_{i}z_{j}}{\left(z_{i}-z_{j}\right)^{2}}\left(2\delta_{\sigma_{i}\sigma_{j}}-1\right),\end{eqnarray}
 where the conserved quantity $M_{\sigma}$ (the number of particles with
spin $\sigma$) is defined as before:\[
M_{\sigma}=\sum_{i}\delta_{\sigma\sigma_{i}}=\int
dx\,\phi_{\sigma}.\]
 The interaction term $\overline{H}_{int}$ can be written
as\begin{eqnarray*}
 \overline{H}_{int} & = & -\beta\sum_{i\ne
j}\theta_{ij}\theta_{ji}\left(-1\right)^{\delta_{\sigma_{i}\sigma_{j}}}\prod_{k\left(\ne
i,j\right)}\left(\frac{z_{k}-z_{i}}{z_{k}-z_{j}}\right)^{\delta_{\sigma_{j}\sigma_{k}}-\delta_{\sigma_{i}\sigma_{k}}}P_{ij}\\
 & = & \frac{\beta}{4}\sum_{i\ne
j}\sin^{-2}\left(\frac{x_{i}-x_{j}}{2}\right)\left(-1\right)^{\delta_{\sigma_{i}\sigma_{j}}}\exp\left\{
\sum_{k\left(\ne
i,j\right)}\left(\delta_{\sigma_{k}\sigma_{i}}-\delta_{\sigma_{k}\sigma_{j}}\right)\ln\left(\frac{z_{k}-z_{j}}{z_{k}-z_{i}}\right)\right\}
P_{ij}.
\end{eqnarray*}
This hamiltonian $\overline{H}$ will now act on bosonic
wavefunctions
 $\chi$. In particular the spin-exchange operator $P_{ij}$ present in
 the interaction term will now act simply as a translation operator, as was seen in earlier sections.

Comparing (\ref{Awata effective hamiltonian}) with (\ref{Haldane
effective hamiltonian}), one can easily see that in order for the
wavefuncion $\chi$ to be bosonic, and the original hamiltonian
fermionic, in both cases ($\tilde{H}$ and $\overline{H}$)
$\Delta^{\beta}$ has to be bosonic. This is due to the factor
$\widetilde{\Delta}$ being antisymmetric in exchange of pairs
$\left(z_{i},\sigma_{i}\right)$.  Note that for $\beta$ even, the
factor $\Delta^{\beta}$ is symmetric under exchange of
$z_{i}\leftrightarrow z_{j}$ (and antisymmetric for $\beta$ odd),
while the factor $\widetilde{\Delta}$ is always antisymmetric under
exchange of pairs $\left(z_{i},\sigma_{i}\right)$.%

As it was done to $H_{\beta+1}$ in section \ref{Collective FT}, we
can obtain the collective form of this hamiltonian (as it is now
bosonic) in coordinate space. Writing each term separately,
$\overline{H}_{1}$ and $\overline{H}_{2}$ are the same as
before:\begin{eqnarray*}
 \overline{H}_{1} & = & -\sum_{\sigma}\int dx\,\phi_{\sigma}\left(x\right)\partial^{2}\Pi^{\sigma}\left(x\right)-\sum_{\sigma}\int dx\phi_{\sigma}\left(x\right)\partial\Pi^{\sigma}\left(x\right)\partial\Pi^{\sigma}\left(x\right)\\
 \overline{H}_{2} & = & \beta\int
dx\phi_{\sigma}\left(x\right)\partial^{2}\Pi^{\sigma}\left(x\right)- \\
  &  & \qquad -\frac{\beta}{2}\sum_{\sigma,\sigma'}\int
dx\phi_{\sigma}\left(x\right)\int
dy\phi_{\sigma'}\left(y\right)\cot\left(\frac{x-y}{2}\right)\left[\partial\Pi^{\sigma}\left(x\right)-\partial\Pi^{\sigma'}\left(y\right)\right].\end{eqnarray*}
 On the other hand we have a new term $\overline{H}_{3}$, which will
be given by:\begin{eqnarray*}  \overline{H}_{3}  & = &
-i\sum_{\sigma}\int dx\int
dy\phi_{\sigma}\left(x\right)\left[\phi_{\sigma}\left(y\right)-\delta\left(x-y\right)\right]\left(1-i\cot\left(\frac{x-y}{2}\right)\right)\partial\Pi^{\sigma}\left(x\right).\end{eqnarray*}
 Finally, we have to deal with the interaction term $\overline{H}_{int}$
and the local term $\overline{H}_{local}$. In fact, once in the
collective form, $\overline{H}_{local}$ can be separated into
$\overline{H}_{local}=\sum_{\sigma}H_{local}^{\sigma}+\sum_{\sigma\ne\sigma'}H_{local}^{\sigma\sigma'}$,
where:\begin{eqnarray*}
 H_{local}^{\sigma} & = & -\frac{\beta}{3}M_{\sigma}\left(M_{\sigma}-1\right)\left(M_{\sigma}-2\right)+\frac{\beta}{2}\left(N-2\right) M_{\sigma}\left(M_{\sigma}-1\right),\\
 H_{local}^{\sigma\sigma'} & = & 2\beta\int dx\int dy\int
dz\frac{e^{iy}}{e^{ix}-e^{iy}}\frac{e^{ix}}{e^{ix}-e^{iz}}\phi_{\sigma}\left(x\right)\phi_{\sigma'}\left(y\right)\left[\phi_{\sigma}\left(z\right)-\delta\left(x-z\right)\right]
\\
 &  & \qquad +\frac{\beta}{4}\int dx\int
dy\phi_{\sigma}\left(x\right)\sin^{-2}\left(\frac{x-y}{2}\right)\phi_{\sigma'}\left(y\right).\end{eqnarray*}
 We conclude that the first term in $\overline{H}_{local}$ is in fact non
local, and contributes directly to the energy.

 By knowing that $P_{ij}$ is applied to a bosonic wavefunction, the collective form of
$\overline{H}_{int}=\sum_{\sigma\ne\sigma'}H_{int}^{\sigma\sigma'}$,
comes out to be\begin{eqnarray*}
 H_{int}^{\sigma\sigma'} & = & -\frac{\beta}{4}\int dx\int dy\frac{\phi_{\sigma}\left(x\right)\phi_{\sigma'}\left(y\right)}{\sin^{2}\left(\frac{x-y}{2}\right)}\exp\left\{ \int dz\ln\left(\frac{e^{iz}-e^{iy}}{e^{iz}-e^{ix}}\right)\left(\phi_{\sigma}\left(z\right)-\delta\left(z-x\right)\right)\right\} \times\\
  &  & \times\exp\left\{ \int dz'\ln\left(\frac{e^{iz'}-e^{ix}}{e^{iz'}-e^{iy}}\right)\left(\phi_{\sigma'}\left(z'\right)-\delta\left(z'-y\right)\right)\right\} \exp\left\{ \int_{y}^{x}dw\partial\left(\Pi^{\sigma}-\Pi^{\sigma'}\right)\right\} .\end{eqnarray*}

We find that the final collective hamiltonian will be given by\[
\overline{H}=\overline{H}_{1}+\overline{H}_{2}+\overline{H}_{3}+\sum_{\sigma\ne\sigma'}\left(H_{local}^{\sigma\sigma'}+H_{int}^{\sigma\sigma'}\right)+\overline{E},\]
 where\begin{equation*}
 \overline{E}  =
\beta^{2}\frac{N\left(N^{2}-1\right)}{12}+\frac{1}{2}\left(\beta+1\right)\sum_{\sigma}M_{\sigma}\left(M_{\sigma}-1\right)+\frac{1-\beta}{3}\sum_{\sigma}M_{\sigma}\left(M_{\sigma}-1\right)\left(M_{\sigma}-2\right).\end{equation*}

 Our objective is to apply the hamiltonian to the bosonic wavefunction $\chi\left(\phi_{\sigma}\right)$.
 The eigenvalue equation gives
$\overline{H}_{\alpha}\left(\chi\right)=E_{\alpha}\chi$, where
$\alpha$ refers to any of the terms of the hamiltonian. We now
determine the various contributions to the energy.

In order to do so we will need the following
results:\begin{eqnarray*}
\qquad\Pi_{\sigma}\left(x\right)\chi & = & \frac{\delta}{\delta\phi_{\sigma}\left(x\right)}\exp\left\{ i\sum_{\sigma'}J_{\sigma'}\int dy\, y\phi^{\sigma'}\left(y\right)\right\} =iJ_{\sigma}x\,\chi,\\
\Pi_{\sigma'}\left(y\right)\Pi_{\sigma}\left(x\right)\chi & = &
iJ_{\sigma}x\,\Pi_{\sigma'}\left(y\right)\chi=-J_{\sigma}J_{\sigma'}xy\,\chi.\end{eqnarray*}
For the first terms of the hamiltonian, the results for the
corresponding energies are straightforward. From $\overline{H}_{1}$
and $\overline{H}_{3}$ terms we have:

\begin{equation}
E_{1}+E_{3}=\sum_{\sigma}J_{\sigma}^{2}M_{\sigma}+\sum_{\sigma}J_{\sigma}M_{\sigma}\left(M_{\sigma}-1\right).\end{equation}
The contribution from the $\overline{H}_{2}$ term is
\begin{equation}
 E_{2} = \beta\sum_{\sigma,\sigma'}\int dx\int
dy\phi_{\sigma}\left(x\right)\phi_{\sigma'}\left(y\right)\frac{e^{ix}J_{\sigma}-e^{iy}J_{\sigma'}}{e^{ix}-e^{iy}}-\beta
N\sum_{\sigma}J_{\sigma}M_{\sigma}\end{equation}

The local term does not have operators, and contributes to the
energy as it is: \[
E_{local}=\sum_{\sigma\ne\sigma'}H_{local}^{\sigma\sigma'}.\]

 According to \cite{ha:1992-}, these local terms will be seen to cancel with local
terms coming from $H_{int}^{\sigma\sigma'}$. We now want to
calculate the energy corresponding to this last term of the
hamiltonian. The corresponding energy (summing over different spins)
is given by \begin{eqnarray}
 E_{int} & = & -\frac{\beta}{2}\sum_{\sigma>\sigma'}\int dx\dashint dy\frac{\phi_{\sigma}\left(x\right)\phi_{\sigma'}\left(y\right)}{\sin^{2}\left(\frac{x-y}{2}\right)}\exp\left\{ \int dz\ln\left(\frac{e^{iz}-e^{iy}}{e^{iz}-e^{ix}}\right)\left(\phi_{\sigma}\left(z\right)-\delta\left(z-x\right)\right)\right\} \times\nonumber\\
  &  & \qquad\times\exp\left\{ \int dz'\ln\left(\frac{e^{iz'}-e^{ix}}{e^{iz'}-e^{iy}}\right)\left(\phi_{\sigma'}\left(z'\right)-\delta\left(z'-y\right)\right)\right\} \left(\frac{e^{ix}}{e^{iy}}\right)^{\left(J_{\sigma}-J_{\sigma'}\right)}.\end{eqnarray}

We are interested in terms up to third order in fields
$\phi_{\sigma}$ and $\Pi_{\sigma}$ (so the $J_{\sigma}$ count as a
field, as they are related to the classical value of the field
$\Pi_{\sigma}$). For that we will basically state the ansatz given
in \cite{ha:1992-} and use their theorems. The main goal is not to
have local terms in the final energy. First, we choose
$M_{\downarrow}\geq M_{\uparrow}$ (no loss of generality). Now to
obtain an energy independent of local terms, we also have to have
$0\leq J_{\sigma}-J_{\sigma'}\leq M_{\sigma'}-M_{\sigma}+1$ for
$\sigma>\sigma'$, and in this case the theorems proved in the same
paper will give:\begin{eqnarray}
 E_{int} & = & -\frac{\beta}{3}\sum_{\sigma>\sigma'}M_{\sigma}\left(M_{\sigma}-1\right)\left(3M_{\sigma'}-M_{\sigma}-1\right)-\beta\sum_{\sigma>\sigma'}M_{\sigma}\left(M_{\sigma'}-M_{\sigma}\right)\left(J_{\sigma}-J_{\sigma'}\right)\nonumber \\
  &  & +\beta\sum_{\sigma>\sigma'}M_{\sigma}\left(J_{\sigma}-J_{\sigma'}\right)^{2}-E_{2}-E_{local}.\end{eqnarray}
 The conclusions we can draw from this result is that if we restrict
$0\leq J_{\sigma}-J_{\sigma'}\leq M_{\sigma'}-M_{\sigma}+1$
($\sigma>\sigma'$), then all contributions from terms of order
$O\left(\left(J_{\sigma}-J_{\sigma'}\right)^{3}\right)$ and higher
vanish. Also the non-local terms will cancel.

The final result for the energy will be, once all terms are gathered\[
E=\sum_{\sigma}E^{\sigma}+\sum_{\sigma>\sigma'}E^{\sigma\sigma'},\]
 where \begin{eqnarray*}
 E^{\sigma} & = & \beta^{2}\frac{N\left(N^{2}-1\right)}{24}+\frac{1}{2}\left\{ \beta N+\left(1-\beta\right)\right\} M_{\sigma}\left(M_{\sigma}-1\right)+\frac{1-\beta}{3}M_{\sigma}\left(M_{\sigma}-1\right)\left(M_{\sigma}-2\right)\\
  &  & +J_{\sigma}M_{\sigma}\left(M_{\sigma}-1\right)+J_{\sigma}^{2}M_{\sigma},\\
 E^{\sigma\sigma'} & = &
-\frac{\beta}{3}M_{\sigma}\left(M_{\sigma}-1\right)\left(3M_{\sigma'}-M_{\sigma}-1\right)-\beta
M_{\sigma}\left(M_{\sigma'}-M_{\sigma}\right)\left(J_{\sigma}-J_{\sigma'}\right)+\beta
M_{\sigma}\left(J_{\sigma}-J_{\sigma'}\right)^{2}.\end{eqnarray*}
 These are the the energies due to 1-spin and 2-spin interactions,
and are the restriction of the general case found in \cite{ha:1992-}
to the $SU\left(2\right)$ case (the sum on $\sigma>\sigma'$ is in
fact just one term, $\sigma=\uparrow,\sigma'=\downarrow$). We have
seen therefore how the nontrivial many-body energy of
\cite{ha:1992-} is obtained in the continuum, collective field
formalism. This example also serves as a demonstration of the method
in describing nontrivial semiclassical backgrounds. Here we had a
state where both the charge and the spin field exhibited nonzero
classical expectation values . We have seen how the collective field
theory is formulated in this new background . The associated
collective field fluctuations give the physical degrees of freedom
in this background.

\section{Conclusions}
We have given a study of continuum, field theoretic techniques of
relevance to matrix and many-body problems. These techniques have
definite condensed matter application. We have featured in the text
various connections  to low dimensional strings. It is expected that
the methods that we have described will be of continuing relevance
in these subjects and will play a role in understanding
nonperturbative physics of low dimensional black holes and
noncritical strings
\cite{Takayanagi:2003sm,Douglas:2003up,Karczmarek:2003pv,McGreevy:2003dn,Das:2004hw,Davis:2004xb,PandoZayas:2005tu}.
This review was concerned with details of the simplest SU(2) theory.
The more general case of SU(R) was given in \cite{Avan:1996vi}. As
we have mentioned of particular interest to string model
applications is the large $R=N$ limit. In this limit one has a full
presence of
 \cite{Jevicki:1996wn} algebra. A basic description of this limit is given in \cite{Jevicki:1996wn}
based on the large N WZW model. Generally this  limit exhibits an
infinite number of bosonic fields and coupling of $W_{\infty}$
degrees of freedom to the original collective boson. Theories of
this kind were given in \cite{Avan:1992gm} and the recent work
\cite{Hatsuda:2006xr}.

The larger Yangian symmetry and  $W_{\infty}$ appearing in these
theories might be of broader relevance for example to
 higher dimensional bosonization \cite{Houghton:2000bn}. There is also the potential
of these models to provide a quantum description of 3 or 4d
noncritical membranes, as was discussed in \cite{Jevicki:1996wn}.
 Finally, the continuum collective  representation (of the
fermion droplet) was seen recently to play a central role in the
 5 dimensional AdS/CFT correspondence through 1/2 BPS states
\cite{Corley:2001zk,Berenstein:2004kk,Lin:2004nb,Iso:1992aa,Donos:2005vm,Donos:2005vs,Agarwal:2006nv}.
It is expected that the continuum field theories of the extended
models will also play a major role in the correspondence, in sectors
with less supersymmetry. The collective field map provided a bridge
between a one-dimensional matrix theory and a 2-dimensional string
theory so it is expected that the extension of this will give a
mapping of the full AdS/CFT correspondence.

\section*{Acknowledgement}
This work was supported in part by the Department of Energy under
Contract DE-FG02-91ER40688, TASK A, and by the fellowship
BD/14351/2003 from \emph{Fundação para a Ciência e Tecnologia},
Portugal. We would like to express our gratitude to the referee for
careful reading of the manuscript and most constructive comments.

\section*{References}
\bibliographystyle{hunsrt}
\bibliography{spin}

\end{document}